\documentclass[a4paper,12pt,english]{article}

\pdfoutput=1

\usepackage[bindingoffset=0.3cm,textheight=22.5cm,hdivide={2.7cm,*,2.7cm}, vdivide={*,22cm,*}]{geometry}

\usepackage{bbold}
\usepackage{authblk}

\usepackage{amsmath,amsfonts,amssymb,babel,slashed,graphicx,color}

\usepackage[numbers,square,comma,sort&compress]{natbib}
\usepackage{adjustbox}
\usepackage[bookmarksnumbered=true,breaklinks=true]{hyperref}

\makeatletter

\global\long\def\bra#1{\left\langle #1\right|}
\global\long\def\ket#1{\left|#1\right\rangle }

\newcommand{\bbm}{\left(\begin{matrix}}
\newcommand{\ebm}{\end{matrix}\right)}
\newcommand{\beq}{\begin{eqnarray}}
\newcommand{\eeq}{\end{eqnarray}}
\makeatother

\newcommand{\del}{\partial}

\newtheorem{prop}{Proposition}[section]

\newtheorem{lemma}[prop]{Lemma}

\newcommand{\be}{\begin{equation}}
\newcommand{\ee}{\end{equation}}

\newcommand{\beqa}{\begin{eqnarray}}
\newcommand{\eeqa}{\end{eqnarray}} \newcommand{\eq}[1]{(\ref{#1})}
\def\nn{\nonumber} \def \bea{\begin{eqnarray}} \def\eea{\end{eqnarray}}

\newcommand{\barr}{\begin{array}}
\newcommand{\earr}{\end{array}}
\numberwithin{equation}{section}

\def\a{\alpha}  
  
 \def\d{\delta} 

 \def\L{\Lambda}

\def\cA{{\cal A}}  \def\cC{{\cal C}} 
   
\def\cG{{\cal G}} \def\cH{{\cal H}}  
\def\cJ{{\cal J}} \def\cK{{\cal K}}  
  \def\cO{{\cal O}} 
\def\cP{{\cal P}} \def\cQ{{\cal Q}}  
\def\cS{{\cal S}}

\def\R{{\mathbb R}} \def\C{{\mathbb C}} 

 \def\one{\mbox{1 \kern-.59em {\rm l}}}

\newcommand{\End}{\mathrm{End}}
\newcommand{\vol}{\rm Vol}
\newcommand{\NC}{{\rm NC}}

\def\bit{\begin{itemize}} \def\eit{\end{itemize}} 
\def\Tr{{\rm Tr}}
\def\tr{{\rm tr}}

\def\({\left(} \def\){\right)}

\def\bra#1{\left\langle#1\right|}
\def\ket#1{\left|#1\right\rangle}
\def\braket#1#2{\left\langle#1|#2\right\rangle}
\def\state#1#2{\big|\mkern-6mu
 {\scriptsize \begin{array}{c}#1\\#2\end{array}}\mkern-6mu
\big)}
\def\statex#1#2{\big({\scriptsize\mkern-6mu\begin{array}{c} #1\\#2\end{array}\mkern-6mu}\big|}

 \allowdisplaybreaks[3]

\makeatother

\renewcommand{\title}[1]{\vspace{10mm}\noindent{\Large{\bf

#1}}\vspace{8mm}} \newcommand{\authors}[1]{\noindent{\large

#1}\vspace{5mm}} \newcommand{\address}[1]{{\itshape #1\vspace{2mm}}}

\begin{document}


 \begin{flushright}
  UWThPh-2022-3 
 \end{flushright}
\begin{center}
\title{ {\Large String modes, propagators and loops on fuzzy spaces} }

\vskip 3mm

\authors{Harold C. Steinacker\footnote{harold.steinacker@univie.ac.at}$\left.^{,\,a}\right.$, Juraj Tekel\footnote{juraj.tekel@fmph.uniba.sk}$\left.^{,\,b}\right.$}

 \vskip 2mm

  \address{ 
$\left.^{\,a}\right.$ {\it Faculty of Physics, University of Vienna\\
 Boltzmanngasse 5, A-1090 Vienna, Austria  }  
 
 $\left.^{\,b}\right.${\it Department of Theoretical Physics,\\ Faculty of Mathematics, Physics and Informatics,\\ Comenius University, Mlynsk\'a Dolina, 842 48 Bratislava, Slovakia }  
   }

\bigskip


\textbf{Abstract}
\vskip 3mm

\begin{minipage}{14cm}%

We present a systematic organization of functions and 
operators on the fuzzy 2-sphere in terms of string modes,
which are optimally localized in position and momentum space.
This allows to separate the semi-classical and the deep quantum regime
of non-commutative quantum field theory, and exhibits its nonlocal nature. 
This organization greatly simplifies the computation 
of loop contributions, avoiding oscillatory integrals and 
providing the effective action directly in position space.
UV/IR mixing is understood as nonlocality arising from long string modes
in the loops. The method is suited for any quantized symplectic space.

\end{minipage}

\end{center}

\tableofcontents


\section{Introduction}

It is expected on general grounds that the incorporation of gravity into a  quantum theory of fundamental interactions will imply some sort of quantum structure of space-time. One approach towards this issue is provided by matrix models, some of which are closely related to string theory \cite{Ishibashi:1996xs,Banks:1996vh}. In this approach, quantum spaces appear naturally as quantized symplectic spaces, which carry a non-commutative gauge theory arising from the fluctuations within the matrix models; see e.g. \cite{Steinacker:2019fcb} for a review and further literature.  Much work has been devoted towards an understanding of these models at the quantum level, which is highly non-trivial due to the UV/IR mixing \cite{Minwalla:1999px}. This phenomenon appears quite universally on any non-commutative geometry and 
leads to novel types of infrared divergences linked to 
ultraviolet divergences in the loops, without any analog in orthodox quantum field theory. 

An appropriate and intuitive understanding of this phenomenon is obtained upon realizing that the UV sector of non-commutative fields is completely non-local, and better described by 
bi-local string modes which have no analog
in conventional field theory. These modes have the structure 
$|x\rangle\langle y|$ where $|x\rangle,|y\rangle$ are coherent states on the quantum space, and behave like open strings connecting $x$ and $y$ \cite{Steinacker:2016nsc,Iso:2000ew}. Most importantly, they are approximate eigenstates of the (matrix) Laplacian which governs the kinematics of the model, 
which allows to compute loop integrals much more efficiently than using the more traditional group-theoretical modes. 
It is then easy to recognize that the UV/IR mixing at one loop is a reflection of a simple non-local term in the effective action, which can be computed in few lines for rather generic 
quantum spaces \cite{Steinacker:2016nsc}. This also exhibits the pathological nature of generic non-commutative field theories in dimensions larger than 2, and the mild form of UV/IR mixing in the  maximally supersymmetrix IKKT model which leads to IIB supergravity interactions in target space \cite{Steinacker:2016nsc}.

In the present paper, we study the properties of string modes in more detail, focusing on the 2-dimensional fuzzy sphere $S^2_N$.
We provide careful justifications and qualifications for the 
approximations which are used in the field-theoretical computations put forward in \cite{Steinacker:2016nsc}. In particular, we derive explicitly the string representation of the propagator, and discuss its properties.
Even though this derivation is specific to the fuzzy sphere, the properties obtained are expected to hold much more generally, and we provide general regularity estimates for string symbols of the propagator.
These results  provide a solid basis for future applications of the method in a more general context. 
The limitation to $S^2_N$ is mainly to simplify the presentation, and most results should generalize in a straightforward way.

The importance of efficient methods for field theory computations 
is hard to overstate. On non-commutative spaces, the only way to proceed so far has been the use of group-theoretical eigenmodes,
i.e. plane waves on $\R^{2n}_\theta$ and (generalized spherical) harmonics on quantized coadjoint orbits.
In the latter case, this makes the computation of any non-trivial
diagrams beyond the most basic ones extremely hard, and completely intransparent. Even on $\R^{2n}_\theta$, the presence of oscillatory integrals in non-planar diagrams is a major problem and also obscures the underlying simple non-local structure. These calculations become much more accessible  using the method of string modes. 
We argue that computations in noncommutative field theory (NCFT) quite generally decompose into semi-classical contributions with an effective UV cutoff provided by the scale of noncommutativity, and a stringy contribution which is easily computable using the novel methods. For sufficiently large mass, the latter regime is dominant, and is analytically accessible.
In particular, we introduce reduced  Feynman rules, which allows to compute and estimate 
loop computations in scalar NCFT in a remarkably simple way. 

There is another motivation for developing these novel tools: to compute the quantum effective action for the 
maximally supersymmetric models related to string theory  \cite{Ishibashi:1996xs,Banks:1996vh}. In these models, the non-local contributions 
due to the long string modes are suppressed by SUSY, and the contributions of the short string modes provide a novel and useful tool to compute e.g. the one-loop effective action on non-trivial backgrounds in position space. Such a computation has been given recently in \cite{Steinacker:2021yxt}, and the present paper provides some further background.

This paper is organized as follows.
After a review of $S^2_N$ emphasizing its distinct regimes, we introduce the string symbol of functions in section \ref{sec:string-IR reg-quant}, and establish  regularity properties in the UV regime. 
In section \ref{sec:op-string} we introduce the string or crossed representation of operators, and derive an explicit formula. This is then applied to the propagator, which
allows to understand its local  and non-local features. A general (off-diagonal) string representation of operators  is also defined, which is shown to be regular and non-oscillatory in the UV.  This is checked numerically for the propagator,
confirming the non-oscillating behavior.
These  results are applied to NCFT in section \ref{sec:loops}.

\section{Functions on $S^2_N$}

Consider the unit sphere $S^2\subset\R^3$ with Cartesian coordinates
\begin{align}
 x_a x^a = 1 \ .
 \end{align}
The fuzzy sphere\footnote{The construction generalizes to any (quantized) 
coadjoint orbit of a compact Lie group, cf.
\cite{Hawkins:1997gj}.} $S^2_N$ 
\cite{hoppe,Madore:1991bw}
is a quantization of $S^2$ with 
 the $SO(3)$-invariant symplectic form $\omega$ (or Poisson structure) satisfying the quantization condition 
 \begin{align}
  \int \omega = 2\pi \dim(\cH)\ .
 \end{align}
Here  $\cH = \C^N$ is the 
irreducible representation of $SU(2)$ with spin
\begin{align}
 \a=\frac{N-1}{2} \ ,
 \end{align}
with generators $J^a_{(N)}$ that satisfy 
\begin{align}
 J^a_{(N)} J_{a(N)}  = \frac 14 (N^2-1) \ =: R_N^2\ .\label{Js}
\end{align}
Then 
the (normalized) fuzzy 2-sphere $S^2_N$ is defined in terms of 
three hermitian matrices 
\begin{align}
 X^a = \frac 1{R_N}\, J^a_{(N)}, 
\end{align}
which satisfy the relations
\begin{align}
[X^a,X^b] = \frac{i}{R_N} \varepsilon^{abc} X^c \qquad  ,
 \qquad  X^a X_a = \one\ .
 \label{fuzzy-S2-def}
\end{align}
Different normalzations are obtained by a trivial rescaling.
The space of (noncommutative)
 functions on $S^2_N$ is given by the operator algebra $\End(\cH)$,
 which decomposes as $SU(2)$-module according to
\begin{align}
 \End(\cH) \cong \cH \otimes \cH^* \cong \bigoplus_{l=0}^{N-1}\ (2l+1)\ .
\end{align}
 Here $(n)$ denotes the $SO(3)$ irrep with dimension $n$. 
 The fuzzy spherical harmonics $\hat Y^l_m$ are defined to be the weight basis of 
 $(2l+1)$.
Explicitly, they are given in terms of the Wigner 3j symbols as
\begin{align}
 \hat Y^l_m=  (-1)^{m} \sqrt{2l+1}\sum_{r,s} \begin{pmatrix}
               l & \a & \a \\
               m & r & -s
              \end{pmatrix}  |r\rangle\langle s| \ 
\label{hat-Y-CG-1}
\end{align}
for $0 \leq l \leq 2\a=N-1$.
Here $|r\rangle, \, r=-\a,...,\a$ is the weight basis of $\cH$
and we have the reality property
\begin{align}
\hat Y^{l\dagger}_m = (-1)^m \hat Y^l_{-m} \ .
\end{align}
The  orthogonality of the 3j symbols
\begin{align}
(2l+1)\sum_{rs}
 \begin{pmatrix}
               l & \a & \a \\
               m & r & -s
              \end{pmatrix}
\begin{pmatrix}
               l' & \a & \a \\
               m' & r & -s
              \end{pmatrix} 
              = \d^{ll'} \d_{mm'}
\end{align}
 implies 
the orthogonality relations and the normalization
\begin{align}
 \tr\big( \hat Y^{l\dagger}_m \hat Y^{l'}_{m'}\big) = \d^{ll'} \d_{mm'}
  = \int_{S^2}dx\,  Y^{l*}_m(x)  Y^{l'}_{m'}(x) \ .
  \label{normalization}
\end{align}
Here and in the following, we denote the usual matrix trace over $\End(\cH)$ by $\tr(\cdot)$ and reserve $\Tr(\cdot)$ for the operator trace over $\End(\End(\cH))$. Also, the integral over functions on $S^2$ is normalized such that $\int_{S^2} = 4\pi$. 
The classical spherical harmonics $Y^l_m(x)$ are normalized accordingly. 
These statements are subsumed in the quantization map 
\begin{align}
 \cQ:\quad L^2(S^2) &\to \End(\cH)  \nn\\
  Y^l_m &\mapsto \left\{\begin{array}{ll}
                         \hat Y^l_m & l < N  \\
                         0 & l \geq N
                        \end{array}
 \right.\ ,
\end{align}
which is unique as an  intertwiner of $SO(3)$ and an isometry w.r.t. the 
inner products defined in \eq{normalization}. 
This map is the analog of the Weyl quantization map on quantum 
mechanical phase space.
 
Finally,  the matrix Laplacian is defined as
 \begin{align}
  \Box \Phi = R_N^2 [X^a,[X_a,\Phi]]\ , \qquad \quad \Phi \in \End(\cH)
  \label{Box-def}
 \end{align}
and it is easy to see that it has the same spectrum $l(l+1)$ for $l=0,1,2,\ldots,N-1$ as the (appropriately rescaled) classical 
Laplacian on the sphere, with eigenfunctions $\hat Y^l_m$.

\subsection{Coherent state representation and symbol for functions}
\label{sec:coh-symbol-funct}

Coherent states on $S^2_N$
are defined as $SU(2)$ orbit of the 
highest weight state $|\a\rangle\in \cH$ \cite{Perelomov:1986tf}.
For any $x\in S^2$ with radius 1, choose some $g_x\in SO(3)$
such that ${x = g_x \cdot p}$, where $p$ is the north pole on $S^2$.
We define 
\begin{align}
 |x\rangle &= g_{x} \cdot |\a\rangle\ , \qquad g_{x} \in \  SU(2)  \nn\\[1ex]
 \langle X^a\rangle &\equiv \langle{ x}| X^a |{ x}\rangle =: {\bf x^a} \ ,
 \qquad  \qquad  {\bf x^a}{\bf x_a} = \frac {(N-1)^2}{N^2-1} =: r_N^2 \ .
\end{align}
Here  $r_N^2$ is the radius of the orbit of coherent states.
These states are in one-to-one correspondence to 
points $x$ on $S^2$ up to a $U(1)$ phase factor\footnote{More precisely, the 
coherent states form a $U(1)$ bundle over $S^2$.}. 
We therefore label them  by $x\in S^2$, where 
the ``north pole`` $p\in S^2$ corresponds to the highest weight state $|\a\rangle$.
Their inner product is given by \cite{Perelomov:1986tf}
\begin{align}
    |\langle x|y\rangle|^2 &= \left(\frac{1+x\cdot y}2\right)^{N-1} =: \frac{4\pi}{N} \d_N(x,y)  \nn\\
 &\approx
    e^{-\frac 12|x-y|_g^2} \ ,\label{def:deltaN}
 \end{align}
where the second line holds for $x,y$ sufficiently close. Therefore 
 \begin{align}
  \langle x|y\rangle &=: e^{-\frac 14|x-y|^2_g}\, e^{i\varphi(x,y)} \ 
 \label{inner-coh}
\end{align}
is exponentially suppressed by
the ''quantum distance`` or metric $|x-y|_g$  defined by 
\begin{align}
 |x-y|_g^2 &:= \frac{|x-y|^2}{L_\NC^2}
 \qquad\quad \mbox{where} \quad |x-y|^2 \equiv \sum\limits_a(x^a - y^a)^2
 \label{inner-coherent}
\end{align}
and characteristic decay length $L_\NC$
\begin{align}
  L_{NC}^2 = \frac{2}{N} \ .
  \label{NC-length}
\end{align}
Here $ \d_N(x,y)$  is a truncation of the standard delta-function, which is optimally localized on $S^2_N$
and normalized such that
\begin{align}
 \int dx\, \d_N(x,y) = 1 \ .\label{delta-normalization}
\end{align}
The  phase factor $e^{i\varphi}$ in \eq{inner-coh}  is gauge-dependent, 
but can be chosen to be the 
symplectic area of the spherical triangle formed by $x,y$, 
and any other reference point.
For example for $x,y$ near the north pole,
we can choose 
\begin{align}
 \varphi = -\frac 12 x^\mu \theta^{-1}_{\mu\nu} y^\nu
 \label{phase}
\end{align} 
where $\theta^{-1}_{\mu\nu}$ is the symplectic form on $S^2$.  
It is easy to see that coherent states are 
optimally localized, i.e. they minimize the uncertainty
\begin{align}
 \Delta^2  
  &= \sum_a\langle (X^a)^2\rangle  - \langle  X^a \rangle^2 
   = 1 - r_N^2  \approx  L_{NC}^2 \ .
 \label{Delta-S4}
\end{align}
Hence $L_\NC$ characterizes the minimal uncertainty on $S^2_N$.
Furthermore, the coherent states $|x\rangle$ on $S^2_N$ form an over-complete basis, with 
\begin{align}
  \one_\cH &= \frac{N}{4\pi}\,\int dx | x\rangle  \langle x| 
   = \frac{\dim\cH}{\vol S^2}\,\int dx | x\rangle  \langle x|  \ .
  \label{coherent-states-S2}
\end{align}
This completeness relation follows easily from $SU(2)$ equivariance and Schur's lemma.

Note that coherent states should {\em not} be interpreted as functions on 
$S^2$: recall that in quantum mechanics, coherent states should be viewed
as functions on position space rather than phase space.
The space of functions on fuzzy $S^2_N$ is given by $\End(\cH)$, which contains
the fuzzy delta-function \eq{fuzzy-delta} 
 built in  terms of coherent states, as a special case of the more general string modes discussed 
in section \ref{sec:string-IR reg-quant}.

\paragraph{Coherent state quantization.}

Using these coherent states, 
we can write down a quantization map 
\begin{align}
 \cQ_c: \quad L^2(S^2) &\to \End(\cH)  \nn\\
  \phi &\mapsto \int  dx \phi(x) | x\rangle  \langle x|
  \label{coh-states-quant}
\end{align}
and a symbol (or de-quantization) map 
\begin{align}
\cS_c: \quad  \End(\cH)  &\to  L^2(S^2)  \nn\\
  \Phi &\mapsto  \langle x| \Phi | x\rangle \ .
  \label{symbol}
\end{align}
It follows immediately from the definitions that these maps are $SO(3)$ 
intertwiners, and moreover $\cQ_c$ is surjective.  
This in turn implies that they map $\hat Y^l_m $ to $Y^l_m $
and vice versa, up to normalization.
We define the normalization constants $c_l$  by 
\begin{align}
 \hat Y^l_m &= c_l \int dx Y^l_m (x) |x\rangle\langle x| \ .
 \label{hatY-Y-relation}
\end{align}
Then
\begin{align}
  Y^l_m(x) &= c_l \langle x| \hat Y^l_m |x\rangle\ ,
  \label{VEV-Y-hat}
\end{align}
thanks to (\ref{normalization})
\begin{align}
 1 = \tr (\hat Y^{l\dagger}_m \hat Y^{l}_{m})
 &= c_l \int dx\,  Y^{l*}_m(x)\langle x| \hat Y^{l}_{m}|x\rangle 
 = \int dx Y^{l*}_m(x) Y^{l}_{m}(x) \ .
\end{align}
Conversely, the symbol of  fuzzy spherical harmonics recovers the classical 
spherical harmonics up to normalization
\begin{align}
  \langle x| \hat Y^l_m |x\rangle=  \frac 1{c_l} Y^l_m(x) \ .
  \label{symbol-Y}
\end{align}
We need to determine $c_l$ explicitly.
As a warm-up, 
$c_0$ is easily obtained from
\begin{align}
 \one &= \frac{N}{4\pi}\int dx |x\rangle\langle x|  , \qquad 
 \hat Y^{0}_0 = c_0 \int dx Y^0_0(x) |x\rangle\langle x|
 \label{normaliz-one}
\end{align}
which using $\hat Y^{0}_0 = \frac{1}{\sqrt{N}}\one$ and $Y^0_0(x) = \frac{1}{\sqrt{4\pi}}$ gives
\begin{align}
 c_0^2 = \frac{N}{4\pi} \ .
 \label{c0-explicit}
\end{align}
$c_1$ is obtained similarly in appendix \ref{sec:asymptotic}.
In general, the  $c_l$ can be obtained as follows: consider
\begin{align}
 \langle \a|\hat Y^{l}_{0}|\a\rangle = \frac 1{c_l} Y^{l}_{0}(p)
  = \frac 1{c_l} \sqrt{\frac{2l+1}{4\pi}} \ ,
\end{align}
where the highest weight state $|\a\rangle$ is located at the north pole $p$.
According to \eq{hat-Y-CG-1}, the 
lhs is nothing but a Clebsch-Gordan coefficient, 
\begin{align}
\langle \a|\hat Y^{l}_{0}|\a\rangle 
 &=  \sqrt{2l+1} \begin{pmatrix}
               l & \a & \a \\
               0 & \a & -\a
              \end{pmatrix}  
 = \sqrt{2 l+1} \sqrt{\frac{((N-1)!)^2}{(N-l-1)! (N+l)!}}
\end{align}
for $n\leq N-1$,
 cf. \cite{nist}.  Therefore
\begin{align}
  c_l &= \frac 1{\sqrt{4\pi}}\sqrt{\frac{(N-l-1)! (N+l)!}{((N-1)!)^2}}
   \ \stackrel{l \ll N}{\sim} \ \sqrt{\frac{N}{4\pi}} \ .
  \label{c-l-result}
\end{align}
For large $N$, we  show\footnote{It is not hard to see that an analogous formula applies to the coherent state 
quantization of plane waves on the Moyal-Weyl quantum plane.} in appendix \ref{sec:asymptotic} that the $c_k$ behave like
\begin{align}
\boxed{\ 
 c_l^2 \ \sim \ \frac{N}{4\pi} e^{N x^2}, 
 \qquad x = \frac{l}{N} \ \in [0,1) \ . \
 }
 \label{c-n-asymptotic}
\end{align}
This suggests to separate  the space of modes on $S^2_N$
into IR  and UV regime as follows
\begin{align}
  \quad l &\leq l_{\rm NC} := \sqrt{N} & & ... \ \mbox{IR regime} &  \nn\\[1ex]
  \quad l &> l_{\rm NC}  & &  ... \  \mbox{UV regime}  &
\end{align}
where $l_\NC$ is the angular momentum corresponding to $L_\NC$.
Therefore $c_l \approx \frac{N}{4\pi} \approx const$ in the IR regime,
where the quantization map and the symbol map are approximately inverse maps:
\begin{align}
 \langle x|\cQ_c(\phi)|x\rangle &\approx \phi, \qquad
 \cQ_c\big(\langle x|\Phi|x\rangle\big) \approx \Phi \ \qquad \mbox{in the IR regime}
 \ . 
\end{align}
This is no longer true in the UV regime where the $c_l$ blows up,
and the symbol (\ref{symbol-Y}) is exponentially suppressed.
The coherent state representation of 
a fuzzy UV mode is thus rapidly oscillating, and  rather misleading. 
This is particularly obvious for  functions 
of the form $\Phi = |x\rangle\langle y|$, which are clearly non-local,
and have a  highly oscillatory  coherent state representation\footnote{Similar remarks apply in quantum mechanics
in the context of the Sudarshan-Glauber representation 
in phase space.
In that context, ''UV regime`` should be replaced by ''deep quantum regime``.}.
A much better representation for such UV modes is provided by string modes, 
which will be discussed below.

\subsection{String modes and semi-classical representation of  functions}
\label{sec:string-IR reg-quant}

The oscillating behavior of the coherent state representation in the UV
can be avoided by adopting  another representation of 
fuzzy functions  via 
bi-local string modes, which are defined as follows  \cite{Steinacker:2016nsc}:
\begin{align}
\state{x}{y}
 &:= \psi_{x,y} := |x\rangle\langle y| \qquad \in \End(\cH)   \nn\\
\statex{x}{y}
 &:= \psi_{x,y}^\dagger := |y\rangle\langle x| \ .
 \label{string-states}
\end{align}
Here $|x\rangle, |y\rangle$ are coherent states\footnote{This construction clearly generalizes to the quasi-coherent states introduced in \cite{Steinacker:2020nva}.}.
String modes are bi-local modes in the algebra of 
functions $\End(\cH)$, with  several remarkable properties. 
They are clearly  bi-local in $x$ and $y$, which is manifest in the inner product 
\begin{align}
 \statex{x}{y}\state{x'}{y'}
  &= \langle x|x'\rangle  \langle y'|y\rangle
  \ \approx \ e^{ - \frac 14 (|x-x'|_g^2 + |y-y'|_g^2)} \ e^{i\varphi} \ ,
  \label{string-states-inner}
\end{align}
using the localization properties of coherent states \eq{inner-coh}. Here $|x-x'|_g$ is the quantum distance, and 
$\varphi=\varphi(x,x';y,y')$ is a gauge-dependent phase.
Moreover, we will see that they are localized in both position and momentum space. 

A first application of string modes is to obtain a representation of any $\cO\in \End(\cH)$ in terms of 
slowly varying functions on $S^2$. This is simply obtained using the completeness relation \eq{coherent-states-S2}, which allows to write any  
$\cO\in \End(\cH)$ in terms of string modes as
\begin{align}
 \cO &= \frac{(\dim\cH)^2}{(\vol S^2)^2} \!\!
 \int\limits_{S^2\times S^2} \!\! \!\! dx dy\, \cO(x,y) |x\rangle\langle y| \ ,
 \label{IR-rep-functions}
 \end{align}
where $\cO(x,y)$ is the {\bf  off-diagonal string symbol}
\begin{align}
 \cO(x,y) := \langle x|\cO|y\rangle \ .
 \label{offdiag-symbol}
\end{align}
Such an off-diagonal representation of $\cO$ is certainly not unique, in view of the diagonal representation \eq{coh-states-quant}. 
However, it is essentially unique if we require $\cO(x,y)$ 
to be {\em in the IR regime},
i.e. the angular momenta are essentially restricted to ${l_x, l_y \leq l_\NC}$. 
We claim that this is indeed the case for \eq{IR-rep-functions}, in contrast to \eq{coh-states-quant}. 
This property will be extremely useful in NCFT, because 
rapid oscillations in the loops are avoided in this way, and all fuzzy functions 
can be described through slowly varying classical functions  $\cO(x,y)$.
Note that the condition 
$l_x, l_y \leq l_\NC$ is not a restriction on the degrees of freedom in any way, it is merely a result of the uncertainly of the quantum space.

A simple consistency check for this claim is obtained by counting the number of modes:
semi-classical functions in two variables comprise roughly 
$(\sqrt{N}^2)^2$ modes, consistent with $\dim (\End(\cH)) = N^2$.
A precise statement is as follows:  let $\cO\in\End(\cH)$ be normalized as 
\begin{align}
 \|\cO\|_{\rm HS}^2 = \Tr(\cO^\dagger\cO) = 1 \ ,
\end{align}
where $\|.\|_{\rm HS}$ denotes the Hilbert-Schmidt norm on $\End(\cH)$.
We claim that $\cO(x,y)$ \eq{offdiag-symbol} is always in the IR regime, i.e. its derivative 
is bounded by the scale of noncommutativity
\begin{align}
\boxed{\
 |\nabla \cO(x,y)| \leq  \frac{1}{L_\NC} \ .
 \ }
\end{align}
To show this,
we first observe that
\begin{align}
  \|\del_\mu|x\rangle\|^2 
   = \big(\del_\mu\langle x|\big) \del_\mu |x\rangle
 &= \frac{\del}{\del x^\mu}\frac{\del}{\del y^\mu}\langle x|y\rangle\big|_{y=0} = \frac{1}{L_\NC^2} \ ,
 \label{del-coherent-bound}
\end{align}
using the explicit form of $\langle x|y\rangle$ \eq{inner-coh},
where $\|.\|$ denotes the norm in $\cH$ and
 $\del_\mu$ denotes a tangential derivative in $T_x S^2$.
Thus the Cauchy-Schwarz inequality implies
\begin{align}
 |\frac{\del}{\del y^\mu} \cO(x,y)|
 = |\langle x|\cO\del_\mu |y\rangle|
 \leq \|\cO^\dagger|x\rangle\| \|\del_\mu|y\rangle\| 
 \leq  \frac{1}{L_\NC}\ ,
 \label{del-strigsymb-est}
\end{align}
since 
\begin{align}
 \|\cO^\dagger|x\rangle\|^2 = \langle x|\cO\cO^\dagger|x\rangle \leq \|\cO\|_{\rm HS}^2 = 1\ ,
\end{align}
using the normalization condition 
${1 = \|\cO\|_{\rm HS}^2 =\tr (\cO^\dagger\cO)}$. 
This bound is essentially saturated by $\cO = |y\rangle\langle y|$,
as it is easy to verify
\begin{align}
  \big|\frac{\del}{\del y^\mu} \langle x|y\rangle \big| \lesssim \frac{1}{L_\NC} \ .
\end{align}
The reason for this mild behavior of the string symbol is of course 
the fact that coherent states are spread over an area $L_\NC^2$, and average out 
any finer oscillations. 
This may seem inconsistent with the fact that the diagonal symbol
faithfully captures the full UV structure, which includes much shorter
wavelengths up to $L_{UV} = \frac 1N = \frac 1{\sqrt{N}} L_\NC$. 
However, this puzzle is resolved by noting that the 
amplitude of such UV modes is strongly suppressed by the factor
$e^{-\frac 12 k^2/N}$ in the string symbol (\ref{c-n-asymptotic}).
In this sense, the extreme UV wavelengths are indeed smoothed out on 
quantum spaces, however at the expense of long-range non-locality
mediated by the string modes.

To proceed,
we need to distinguish between a semi-classical and a deep quantum regime of the string modes:
The {\bf short or local string modes}  $\state{x}{y}$
for $|x-y| \leq L_\NC$ provide the noncommutative analog of optimally localized 
wave packets, with characteristic size $L_\NC$
and linear momentum determined by $x-y$.
On the other hand,  the {\bf long string modes} for $|x-y| > L_\NC$
are completely non-local and non-commutative.
They provide the appropriate description of the UV or deep quantum 
regime of NCFT.

\subsection{Short string modes as localized wave-packets on $S^2$}

Consider first the semi-classical regime and its description in terms of 
short string modes.
We claim that the following identification
\begin{align}
 \boxed{\
\state{y+\frac k2}{y-\frac k2} \ \cong \ \psi_{\tilde k;y}} \ , \ 
 \qquad  \tilde k_\mu = k^\nu\theta^{-1}_{\nu\mu}  = \frac{1}{L^2_\NC}k^\nu\epsilon_{\nu\mu} 
 \label{string-wavepacket-identif}
 \end{align}
 
 on $T_y S^2 \cong \R^2$
 defines an isometry from short string modes on $S^2_N$ to the following classical 
 wave packets  near $y$
  \begin{align}
  \psi_{\tilde k;y}(x) &= \sqrt{\frac{2}{\pi L_\NC^2}}\,  e^{-\frac i2 \tilde k y}  e^{i \tilde k x} e^{-|x-y|_g^2} \ .
  \label{wavepacket}
\end{align}
These wave-packets have linear momentum $k$ and characteristic size $L_\NC$.
To justify this identification\footnote{On the Moyal-Weyl quantum plane $\R^2_\theta$, this identification can be obtained from the Wigner map.}, we check that the inner products agree. 
A simple Gaussian integration gives
 \begin{align} 
 \langle \psi_{\tilde k;a},\psi_{\tilde l;b} \rangle 
 &= e^{\frac i2 (\tilde l - \tilde k) (a+b) } e^{ -\frac 12|a-b|_g^2 -\frac{1}8 |k-l|_g^2} \ ,
 \label{inner-prod-wavepacket}
\end{align} 
which is consistent with the inner product \eq{string-states-inner} for the short string modes 
\begin{align}
  \statex{a+\frac k2}{a-\frac k2}\state{b+\frac l2}{b-\frac l2} 
  &\approx \ e^{ - \frac 14 |a-b + \frac 12(k-l)|_g^2 -  \frac 14 |a-b-\frac 12(k -l)|_g^2} \ e^{i\varphi} \nn\\
  &= e^{-\frac 12|a-b|_g^2 - \frac 18 |k-l|_g^2}\, e^{i\varphi} \ .
 \label{inner-prod-string-short}
\end{align}
The phase factor is recovered in the gauge \eq{phase} where $\varphi$ is the difference of 
the symplectic triangles $(a+\frac k2,b+\frac l2,0)$
and $(a-\frac k2,b-\frac l2,0)$:
\begin{align}
 \varphi = \frac 12(a + \frac k2)(\tilde b + \frac {\tilde l}2) 
 - \frac 12(a - \frac k2)(\tilde b - \frac {\tilde l}2)
 = \frac 12(a \tilde l - b \tilde k) \ .
\end{align}
Taking into account the phase factor
$e^{\frac i2 \tilde k a}$ in \eq{wavepacket},  the phase of the inner product becomes 
\begin{align}
 \varphi = \frac 12(a \tilde l - b \tilde k - a \tilde k + b \tilde l)
   = \frac 12 (a+b) (\tilde l- \tilde k) \ ,
\end{align}
consistent with \eq{inner-prod-wavepacket}. 
The momentum assignment  will be clarified further
in section \ref{sec:kinematics-strings}.
However, the symbol of the short string modes
\begin{align}
\statex{x}{x}
 \state{y+\frac k2}{y-\frac k2}
 = e^{-|x-y|^2_g }e^{-\frac 14 |k|^2_g }
\end{align}
differs slightly from the isometric identification \eq{string-wavepacket-identif}.
This indicates that the short string states are outside of (but bordering on) the 
semi-classical regime, which 
is reflected in  peculiar algebraic properties of the string modes such as
\begin{align}
|x\rangle\langle y||y\rangle\langle z| = |x\rangle\langle z| \ 
\end{align}
which are not reproduced by their symbols\footnote{Of course this could be reconciled in terms of a star product  on the classical space of functions.
Here we simply wish to point out that they are not quite in the semi-classical regime.}.
The reason is that the size of the wave-packet coincides with the uncertainty scale $L_\NC$.
Semi-classical wavefunctions should accordingly be realized
as superpositions of short string states such as
 \begin{align}
  \hat\psi^{(L)}_{\tilde k;y} & \sim
   \int\limits_{S^2} dz\, e^{-|y-z|^2/L^2}\state{z+\frac k2}{z-\frac k2}\ ,
 \end{align}
corresponding to a Gaussian wave packet of size $L$ centered at $y$ 
with momentum $\tilde k$.
On the other hand, the short string modes with $k=0$ 
can be identified with fuzzy delta-functions.

\paragraph{Fuzzy delta-function as short string mode.}

Consider the string mode 
\begin{align}
 \hat \d_p :=\left|^p_{p} \right) = \sum_{l} a_l\, \hat Y^l_0 
 \label{fuzzy-delta}
\end{align}
at the north pole $p\in S^2$. According to the above discussion, this should be 
interpreted as optimally localized wavepacket at $p$ without linear momentum,
i.e. as fuzzy delta-function at $p$.
Indeed, 
\begin{align}
 \tr\big(\hat \d_p \hat Y^l_m \big) = \langle p| \hat Y^l_m |p\rangle
  = \frac{1}{ c_l} Y^l_m(p) 
\end{align}
using \eq{VEV-Y-hat}, and similarly for any point on $S^2$.
The coefficients $a_l$ are obtained using   \eq{normalization} 
\begin{align}
 a_l &=  \langle p|\hat Y^l_0|p\rangle 
=   \frac{1}{c_l}\, Y^l_0(p) 
 =  \frac{1}{c_l}\,\sqrt{\frac{2l+1}{4\pi}} 
 \ \sim \ \sqrt{\frac{2l+1}{ N}} e^{-\frac{l^2}{2N}} \ .
\end{align}
As a consistency check, we compute
\begin{align}
1 &=  \tr(\hat \d_p\hat \d_p ) = 
 \sum_l a_l^2 = \sum_{l=0}^{N-1} \frac{2l+1}{4\pi c_l^2}
 \sim \frac{2N^2}{4\pi}\int_0^{1}dx \frac{x}{c_x^2} \nn\\
 &= 2N\int_0^{1}dx\, x e^{-N x^2} 
  = 1\ ,
\end{align}
using the asymptotic formula \eq{c-n-asymptotic}.
Note that the function $e^{-N x^2}$ 
provides a cutoff at $x\sim \frac{1}{\sqrt{N}}$, 
so that  only modes in the IR sector $k\sim\sqrt{N}$ contribute, as expected.
In particular, the optimally localized fuzzy 
(delta-like)  function at the north pole is 
\begin{align}
  \hat \d_p = \left|^p_{p} \right) 
 &= \sum_{l=0}^{N-1} \,\sqrt{\frac{2l+1}{N}}e^{-\frac{l^2}{2N}} \hat Y^l_0 
 \ \approx \ \sum_{l=0}^{\sqrt{N}} \sqrt{\frac{2l+1}{N}} \, \hat Y^l_0 \ ,
 \label{delta-NC}
\end{align}
with classical counterpart
\begin{align}
 \delta_{p,N}(x) &:= \ \sum_{l=0}^{\sqrt{N}} \sqrt{\frac{2l+1}{N}} \,  Y^l_0(\vartheta,\varphi)
 \approx  \sqrt{N} e^{-N x^2} \ .
 \end{align}
This is indeed the truncation of the exact delta-function on $S^2$
at ${l=l_\NC=\sqrt{N}}$.

\paragraph{Fundamental solution on $S^2_N$.}

In particular, the fuzzy analog of the fundamental solution of $(\Box+m^2)$
centered at the north pole is
\begin{align}
 \cG_p &= \frac 1{\Box + m^2} \left|^p_{p} \right) 
 := \frac{1}{\sqrt{N}}\sum_{l=0}^{\sqrt{N}}  \frac{\sqrt{2l+1}}{l(l+1) + m^2}
  \hat Y^l_0 \ ,
  \label{fund-G-def}
 \end{align} 
 which satisfies
 \begin{align}
 (\Box+m^2)\cG_p &=  \hat \d_p \ \qquad \in \End(\cH) \ .
 \end{align}
 We can construct the corresponding fuzzy 2-point function $ G_N(x,y)$ as follows. We first compute
\begin{align}
 G_N(x,p) :=
   \left(^x_{x} \right| \frac 1{\Box + m^2} \left|^p_{p} \right) 
 &= \frac{1}{\sqrt{N}}\sum_{l=0}^{\sqrt{N}}  \frac{\sqrt{2l+1}}{l(l+1) + m^2}
 \frac 1{c_l} Y^l_0(x) \nn\\
   &= \frac{1}{N}\sum_{l=0}^{\sqrt{N}}  \frac{2l+1}{l(l+1) + m^2}
   P_l(\cos\vartheta)\ .
 \end{align}
Thanks to the symmetry of the sphere, the general formula will have the same form, with $\vartheta$ understood as the angle between the points $x$ and $y$, rather than the azimuthal coordinate of $x$. Thus
\begin{align}
 G_N(x,y) :=
   \left(^x_{x} \right| \frac 1{\Box + m^2} \left|^y_{y} \right) 
 &=\frac{1}{N}\sum_{l=0}^{\sqrt{N}}  \frac{2l+1}{l(l+1) + m^2}
   P_l(\cos\vartheta) \ \nn\\
  &\approx \frac 1{N}\ln(|x-y|^2) \ , \qquad m^2 \ll l^2_\NC
   \label{NC-Greens-1}
 \end{align}
which  is a regularization of the fundamental solution
for distances greater than $L_{NC}$.
This becomes more transparent
noting that  
  $ \statex{x}{x}$ amounts to the equivariant symbol map \eq{symbol}, so that
\begin{align}
(\Box_x+ m^2)G_N(x,y) =
 (\Box_x+ m^2)\left(^x_{x} \right|\frac 1{\Box+m^2} \left|^y_{y} \right)
  &=  \statex{x}{x} \state{y}{y} = \frac{4\pi}{N} \d_N(x,y)\ .\label{fuzzy-2point-def}
\end{align}
 In particular, we obtain
 \begin{align}
  \left(^p_{p} \right|\frac 1{\Box+m^2} \left|^p_{p} \right)
 &\approx \frac 1N \sum_{l=0}^{\sqrt{N}} \frac{2l+1}{l(l+1)+m^2} 
 \sim \frac 1N\ln(N)
 \label{delta-prop-1}
\end{align}
as long as $m^2 \ll l^2_\NC$, consistent with \eq{NC-Greens-1} for 
$|x-y| \approx L_\NC$.

Note that the large $N$ limits of the fuzzy expressions like \eq{NC-Greens-1} or \eq{fuzzy-2point-def} differ from the their commutative counterparts by a factor of $\frac{4\pi}{N}$ due to the presence of this factor in the coefficients $c_l$.

\subsection{Kinematical properties of string modes}
\label{sec:kinematics-strings}

In the context of matrix models, differential operators  are realized in terms of 
commutators i.e. derivations. 
For the fuzzy sphere, consider
the following derivative operators acting on $\End(\cH)$:
\begin{align}
 \cP^a\, \Phi &:= [X^a, \Phi], \qquad \Phi \in \End(\cH)   \nn\\[1ex]
 \Box\, \Phi  &:=  R_N^2\, \cP^a \cP_a \Phi \ ,
 \label{P-P2-NC}
\end{align}
where the factor $R_N^2$ is inserted for consistency with \eq{Box-def}.
These can be viewed as quantized differential operators on $S^2$.
If $\Phi$ is a (quantized) function on $S^2$ in the semi-classical regime
i.e. with wavelength greater than $L_{\rm NC}$, then $\cP^a \sim i\{x^a,.\}$ is adequately interpreted as derivative operator:
\begin{align}
  \cP^a\, \Phi &= [X^a, \Phi] = i\theta^{a\mu} \del_\mu \Phi, 
  \qquad \theta^{a\mu} = \{x^a,y^\mu\}
\end{align}
in local coordinates $y^\mu$.
However if $\Phi$ is outside of the semi-classical regime,
$\cP^a \in \End(\End(\cH))$ 
should be viewed as a {\em non-local} operator.
This non-locality  plays a crucial role in NCFT, due to virtual modes propagating in the loops. It is manifest 
by considering their matrix elements w.r.t. string modes 
\begin{align}
\statex{x}{y}  \cP^a \state{x}{y}
 = {\bf x^a}(x) -  {\bf x^a}(y)
 \approx x^a - y^a \ .
 \label{Pa-exp}
 \end{align}
Hence the string modes $\state{x}{y}$ have ``matrix momentum'' 
$\cP = x -   y$, which confirms the 
identifcation \eq{string-wavepacket-identif} with semi-classical 
wavepackets on the tangent space $\R^2_\theta$ with wave-number $\tilde x - \tilde y$.
The general matrix elements
\begin{align}
\statex{x}{y}  \cP^a \state{x'}{y'}
 &\approx ({\bf x^a}(x) -  {\bf x^a}(y)) 
 \langle x | x'\rangle \langle y' | y\rangle 
 \end{align}
are approximately diagonal.
This confirms previous observations \cite{Bigatti:1999iz,Bergman:2000cw,Jiang:2001qa} in noncommutative field theory,
which now acquire a precise mathematical realization.

Similarly for the Laplacian, we obtain
\begin{align}
R_N^{-2}
\statex{x}{y} \Box \state{x'}{y'}
 = \statex{x}{y} \cP^a \cP_a \state{x'}{y'}
 &= \langle x|X^a X_a |x'\rangle \langle y' | y\rangle  + \langle x | x'\rangle \langle y'| X^a X_a |y\rangle 
 - 2  \langle x| X^a |x'\rangle \langle y'| X_a |y\rangle \nn\\
  &\approx  E_{xy} \, \langle x | x'\rangle \langle y' | y\rangle \ 
  \label{propagator-general}
\end{align}
to a very good approximation, where
 \begin{align}
 E_{xy} &= (\vec{\bf x}(x) -\vec{\bf x}(y))^2 +  2 L_\NC^2  
 \label{PP-strings}
\end{align}
is the energy of a string mode, which is  given by its length square
plus the intrinsic quantum length scale.

\paragraph{The propagator.}

Since the Laplacian is almost diagonal on the string modes, we can expect the following approximate formula for the propagator 
in the string basis
\begin{align}
 \statex{x}{y}(\Box + m^2)^{-1}\state{x'}{y'}
  \approx \left\{\begin{array}{ll}
                 G_\L(x,x')   & \quad  x=y \neq x'=y'   \\[1ex]
                \tilde G(x,y) &\quad  x=x' \neq  y=y' \\[1ex]
                  0  & \quad\mbox{otherwise}
                \end{array}  \right. \ .
\label{string-propagator-symbol}
\end{align}
Here 
\begin{align}
 G_\L(x,x') &:= \statex{x}{x}(\Box + m^2)^{-1}\state{x'}{x'} \approx \frac{4\pi}{N} G(x,x')  
 \nn\\
 (\Box + m^2)G_\L(x,x') &= \statex{x}{x}\state{x'}{x'}  = \frac{4\pi}{N}  \d_N(x,x')
  \label{prop-class-approx} 
 \end{align} 
  is a regularization of the {\em commutative} Greens function $G(x,y)$
for $\Box+m^2$ with UV cutoff $\L$, while
\begin{align}  
  \tilde G(x,y) &:= \statex{x}{y}(\Box + m^2)^{-1}\state{x}{y}
 \approx \frac{1}{R_N^2|x-y|^2 + m^2 } \ 
 \label{prop-NC-approx}
\end{align}
is the UV contribution to the propagator which arises from the string modes (\ref{string-symbol-large-N}). 
Note that both are well-defined function, in contrast to the classical Greens function $G(x,y)$  which is a distribution.
These two functions should merge at coincident points, 
\begin{align}
 G_\L(x,x) \approx \tilde G(x,x)
    \approx \frac{1}{ m^2 + N}  \ .
\end{align}
This is  consistent with \eq{delta-prop-1} 
up to a factor $\ln(N)$, which is missed by the  string approximation.
This supports the validity of the approximation \eq{prop-NC-approx}
 in the nonlocal regime $|x-y| > L_{NC}$.
The following section is devoted to refinement and justification the string representation
of the propagator.

\section{Operators on $S^2_N$ and string representations}
\label{sec:op-string}

 In analogy to the coherent state realization of functions ${\Phi\in\End(\cH)}$
  discussed above, 
 operators $\cO\in \End(\End(\cH))$
on fuzzy spaces -- such as the propagator -- can be realized in terms of  string modes. 
 However, there are now two very different possibilities: First, there is a  ''local`` realization in terms of local string modes,
 which is analogous to the standard representation of operators via an 
 integral kernel. 
 Second, there is an entirely new representation
 in terms of non-local string modes, which can be viewed as a crossed version 
 of the local one. This version has no classical analog, and it will  
 be extremely useful for the propagator and loop computations in NCFT.
 This is the main focus of the present paper.
 
 Moreover in both cases, one should distinguish  between 
 a minimal, ''diagonal`` realization  which is in general highly oscillatory,
 and an off-diagonal representation  
 which is typically much better behaved, notably for the propagator.

\subsection{Operator kernel and local representation}
\label{sec:coh-kernel-ops}

We start with the representation of operators which corresponds to the classical 
integral kernel of operators.
Consider an operator  on fuzzy $S^2_N$
\begin{align}
 \cO: \quad \End(\cH) \to \End(\cH)\ ,
\end{align}
or equivalently ${\cO\in \End(\End(\cH))}$.
We can decompose $\End(\End(\cH))$ as $SO(3)_L \times SO(3)_R$ module as
\begin{align}
 \End(\End(\cH)) &\cong  \End(\cH_L) \otimes \End(\cH_R)^* \nn\\
&\cong ((N)_L\otimes (N^*)_L) \otimes ((N)_R\otimes (N)_R^*) \nn\\
&\cong (\oplus_l (l)_L) \otimes (\oplus_l (l)_R)
\label{End-End-rep}
\end{align}
into irreps of $SO(3)_L \times SO(3)_R$. This is isomorphic to the decomposition of 
the space of functions on $S^2_L\times S^2_R$ under $SO(3)_L \times SO(3)_R$ 
\begin{align}
 L^2(S^2_L \times S^2_R )
 &\cong (\oplus_l (l)_L) \otimes (\oplus_l (l)_R) \ ,
\end{align}
truncated at $l_{\rm max}$. Therefore there is a unique intertwiner which respects the normalization given by the integral and trace, respectively.
Relaxing the isometry requirement, we can define such an intertwiner  
analogous to the coherent state quantization  \eq{coh-states-quant} as
follows 
\begin{align}
\boxed{\ 
\begin{aligned}
 \cK: \quad L^2(S_L^2\times S_R^2) &\to End(End(\cH))  \\
   k(x,y) &\mapsto \int\limits_{S^2\times S^2} dx dy \state{x}{x}
 k(x,y) \statex{y}{y} \ .
 \end{aligned} 
 \ }
 \label{quant-End-kernel}
\end{align}
This map is surjective because \eq{coh-states-quant} is,
and we denote the function $k(x,y)$  as 
{\em local kernel} of $\cO$.
Since $\state{x}{x}$ is the noncommutative analog of a delta-function,
 $k(x,y)$ is analogous to the standard integral kernel of an operator.
Conversely,  consider the 
following expectation value of operators for local string modes
at different locations:
\begin{align}
\boxed{ \
 \cO_N(x,y) :=  \statex{x}{x} \cO \state{y}{y} \ .
 \ }
\end{align}
In view of the inner products \eq{string-states-inner},
this provides an approximate inverse to the kernel map \eq{quant-End-kernel},
\begin{align}
  \cK\big(\cO_N(x,y)\big) \approx \cO\ , 
  \qquad \big(\cK(k(x,y))\big)_N \approx k(x,y)
  \label{op-kernel-symbol}
\end{align}
provided (!) the functions are sufficiently smooth, i.e. not oscillating at scales 
shorter than $L_{\rm NC}$. Otherwise, this representation of $\cO$ is 
exact but misleading.
 For example, the propagator is  represented by the kernel $k(x,y)$ as follows
\begin{align}
 (\Box+m^2)^{-1} &= \sum \frac 1{l(l+1)+m^2} \hat Y^{l}_{m} \otimes \hat Y^{l\dagger}_{m} \nn\\
  &= \cK\Big(\sum \frac 1{l(l+1)+m^2} c_l^2 Y^{l}_{m}(x) Y^{l*}_{m}(y) \Big)\nn\\
  &= \cK\Big(\sum \frac 1{l(l+1)+m^2} \frac{2l+1}{4\pi} c_l^2 P_l(\cos\vartheta)\Big) \nn\\
  &=: \cK(k(x,y))\ ,
\end{align}
where $x\cdot y = \cos\vartheta$,
using the spherical harmonics addition theorem \eq{spher-add-thm}.
Even though this formula is exact, there is a problem, because
 the $c_l$ blow up for 
$l\gg l_\NC$ \eq{c-n-asymptotic}. Therefore  $k(x,y)$ is highly 
oscillatory, and  very different from its classical cousin
\begin{align}
 G(x,y) = \sum\limits_{l=0}^\infty \frac 1{l(l+1)+m^2} \frac{2l+1}{4\pi} P_l(\cos\vartheta)\ ,
\end{align}
which is the fundamental solution of
 \begin{align}
 (\Box_x +m^2)G(x,y) &=  \d_y(x) \   .
 \end{align}
 In contrast,
the desired UV truncation 
is recovered by the local string symbol:
\begin{align}
 G_N(x,y) &:=  \statex{x}{x}\frac 1{\Box+m^2} \state{y}{y} \nn\\
  &= \sum_l \frac 1{l(l+1)+m^2} \frac 1{c_l^2} Y^{l}_{m}(x) Y^{l*}_{m}(y) \nn\\
  &\approx \frac 1N\sum\limits_{l=0}^{l_\NC} \frac {2l+1}{l(l+1)+m^2}  P_l(\cos\vartheta)  \ 
 \label{2-point-recovery}
\end{align}
for $m^2 \ll l_\NC^2$ in agreement with \eq{NC-Greens-1},
using \eq{VEV-Y-hat} and the spherical harmonics addition theorem \eq{spher-add-thm}.
We will see that the symbols are typically better behaved in the UV
than the operator kernel.
As noted in the previous section, the large $N$ limit of fuzzy fundamental solution $G_N(x,y)$ differs from the classical $G(x,y)$ by a factor
\begin{align}
    G_N(x,y)\to \frac{4\pi}{N} G(x,y)\ .
\end{align}

The above example of the propagator illustrates the non-classical nature of  
the local operator kernel in the UV regime: 
Even though this representation always exists, it may be highly 
singular. This is reminiscent of the 
coherent state quantization map discussed in section \ref{sec:coh-symbol-funct}, and
we will encounter 
a similar phenomenon in section \ref{sec:string-symbol-ops}. This singular behavior will be resolved in section \ref{sec:string-IR-op},  similarly to the 
IR representation of functions in \eq{IR-rep-functions}.

\subsection{String kernel, string representation and crossing}
\label{sec:string-rep-ops}

Now we discuss a different representation of operators in terms of 
non-local string modes, which is particularly useful in the UV regime.
This is a crossed version of the representation discussed in the previous 
section. For some operators such as the propagator, this 
turns out again to be singular, which will be cured in section \ref{sec:string-IR-op}.

The key to the string representation of operators is to consider a different 
action of $SO(3)_L \times SO(3)_R$ in the identification \eq{End-End-rep}:
\begin{align}
 \End(\End(\cH)) &\cong \End(\cH) \otimes \End(\cH)^* \nn\\
&\cong ((N)_L\otimes (N^*)_R) \otimes ((N)_R\otimes (N)_L^*)\ ,
\label{End-crossed}
\end{align}
where $SO(3)_L \times SO(3)_R$ acts as indicated.
Using the coherent state representation \eq{coh-states-quant} of $\End(\cH)$, we can again identify this decomposition 
with functions on $S^2 \times S^2$ via the following intertwiner:
\begin{align}
 \tilde\cQ: \quad \cC(S_L^2\times S_R^2) &\to \End(\End(\cH))  \nn\\
   s(x,y) &\mapsto \int\limits_{S^2\times S^2} dx dy \state{x}{y}
 s(x,y) \statex{x}{y}  \ .
 \label{quant-End-End}
\end{align}
By construction, this respects the (modified) action of $SO(3)_L \times SO(3)_R$,
and the map is surjective because \eq{coh-states-quant} is.
We denote $s(x,y)$ as {\em string kernel} of the operator.
Explicitly, the map is given by
\begin{align}
 \tilde\cQ\big( Y^l_m(x)  Y^{l'}_{m'}(y)\big) 
  &= \int dx Y^l_m (x) |x\rangle\langle y|\otimes
  \int dy Y^{l'}_{m'} |y\rangle\langle x| = \cC(c_l^{-1} c_{l'}^{-1} \hat Y^l_m \otimes \hat Y^{l'}_{m'})= \nn\\
   &= \frac 1{c_l c_{l'}} \sqrt{(2l+1)(2l'+1)}
   \begin{pmatrix}
               l & \a & \a \\
               m & r & -s
              \end{pmatrix}
              \begin{pmatrix}
               l' & \a & \a \\
               m' & r' & -s'
              \end{pmatrix}
              |r\rangle\langle s'|\otimes |r'\rangle\langle s|\ .
\label{cQ-map-crossed}
\end{align}
where $\cC$ denotes the crossing of the tensor factors
$2 \leftrightarrow 4$,
\begin{align}
\cC:\quad \End(\cH) \otimes \End(\cH)^* &\cong \End(\cH) \otimes \End(\cH)^*  \nn\\[1ex]
 (N)_L\otimes  (N)_L^* \otimes (N)_R\otimes (N^*)_R
 &\mapsto (N)_L\otimes (N^*)_R \otimes (N)_R\otimes (N)_L^* \ .
\label{C-crossed}
\end{align}
The advantage of this representation -- which is possible only in the fuzzy case -- is that e.g.
the Laplacian is represented in a very simple way, making 
manifest the fact that the string modes are approximate eigenstates \eq{propagator-general}.
This is very useful in the 
context of noncommutative field theory.

As for the operator kernel \eq{op-kernel-symbol}, the string kernel of an operator $\cO$ can be recovered approximately from the 
{\em string symbol} of $\cO$, which we define as follows
\begin{align}
O(x,y) := \statex{x}{y} \cO \state{x}{y}  &\approx s(x,y)\  ,
\label{string-symbol}
\end{align}
which is an intertwiner of the above $SO(3)_L \times SO(3)_R$ action.
The approximation is good 
provided all functions are in the IR regime.

In particular, consider {\em invariant operators} or functions, 
which by definition 
are invariant under $SO(3)_{\rm diag} \subset SO(3)_L \times SO(3)_R$. 
This 
concept coincides for the local and the string representation.
Then the symbols are functions ${s(x,y) = s(gx,gy)}$ which are invariant under $g\in SO(3)$, which implies that ${s(x,y) = s(|x-y|)}$. Therefore invariant operators can be written 
in terms of just one function in two different ways, as
\begin{align}
\cO =  \int dx dy \state{x}{y}
 s(|x-y|) \statex{x}{y}
  =   \int dx dy \state{x}{x}
 k(|x-y|) \statex{y}{y} \ .
 \label{inv-op-symbol}
\end{align}
The string kernel $s(|x-y|)$  can be obtained from the standard kernel $k(|x-y|)$ using a crossing relation, 
which for the present case of $S^2$ is closely related to $6j$ symbols.
For invariant operaors $\cO$,
this is obtained using the following result:
\begin{lemma}
\label{lem:cross}
\begin{align}
\sum_m\hat Y^l_m \otimes \hat Y^{l\dagger}_{m}
&= \sum_{k} A^{l k} \cC(\sum_n\hat Y^k_n \otimes \hat Y^{k\dagger}_{n})\ ,
\end{align}
where
\begin{align}
 A^{lk} &=  \sum_k (2l+1) (-1)^{l+k+2\a} \left\{
\begin{array}{lll}
 k & \a & \a \\
 l & \a & \a
         \end{array}
     \right\} 
 \ \approx\ \sum_k \frac{2l+1}{2\a}P_k(1-\frac{l^2}{2\a^2})
    \label{Alk-def}
\end{align}
satisfies $A^2 = \mathbb{1}$. The approximation holds if either 
$l\ll\a$ or $k\ll\a$.
\end{lemma}
The proof is given in appendix \ref{sec:app-lemma}.

Applied to  \eq{cQ-map-crossed}, this  gives
\begin{align}
\tilde\cQ\big(\sum_m Y^l_m(x) Y^{l*}_{m}(y)\big)
 &= \frac 1{c_l^2} \cC\big(\sum_m \hat Y^l_m \otimes \hat Y^{l\dagger}_{m}\big)  
 = \frac 1{c_l^2} \sum_k A^{lk}\sum_n \hat Y^k_n \otimes \hat Y^{k\dagger}_{n} \nn\\
 \tilde\cQ\big(\sum_l A^{kl} c_l^{2}\sum_m Y^l_m(x) Y^{l*}_{m}(y)\big)
 &= \sum_n \hat Y^k_n \otimes \hat Y^{k\dagger}_{n}\ .
\end{align}
Together with the spherical harmonics addition theorem \eq{spher-add-thm}, 
we obtain the string kernel of any function of the Laplacian as
\begin{align}
f(\Box) &= 
 \sum_{k,m}f(k(k+1))\hat Y^k_m \otimes \hat Y^{k\dagger}_{m}  =  \tilde\cQ\Big(\sum_{l,k} f(k(k+1)) A^{kl} c_l^{2}
 \frac{2l+1}{4\pi} P_l(\cos\vartheta)\Big) \nn\\
 &= \tilde\cQ\Big(\sum_{k,l}(2k+1) (-1)^{l+k+2\a} f(k(k+1))
  c_l^2 \left\{
\begin{array}{lll}
 l & \a & \a \\
 k & \a & \a
         \end{array}
     \right\} \frac{2l+1}{4\pi} P_l(\cos\vartheta) \Big)  \nn\\
 &=: \tilde\cQ\big(F(|x-y|)\big) \ .
 \label{Propagator-formula}
\end{align}
Hence the desired string  representation for the propagator is given by 
\begin{align}
 f(\Box) = \int dx dy\, \state{x}{y}
 F(|x-y|)
     \statex{x}{y}\ ,
\end{align}
with $F$ defined in \eq{Propagator-formula}.
However, since $c_l$ grows exponentially 
\eq{c-n-asymptotic}, this form is only useful if the sum terminates after a few terms, otherwise it exhibits a rapidly oscillating behavior arising from the 6j symbol for large $l$. It works well for the Laplacian and 
finite powers thereof, but not for the propagator.
The off-diagonal IR representation discussed in section \ref{sec:string-symbol-ops}
is much better suited for the propagator.

The above  formula can be evaluated and checked explicitly 
for the following special cases:

\paragraph{The identity operator $\one$.}

In this case, \eq{Propagator-formula} gives
\begin{align}
 \one &= \int\limits dx dy \state{x}{y}\statex{x}{y}
    \sum_{k,l}(2k+1) (-1)^{l+k+2\a} c_l^2 \left\{
\begin{array}{lll}
 l & \a & \a \\
 k & \a & \a
         \end{array}
     \right\} \frac{2l+1}{4\pi} P_l(\cos\vartheta) \nn\\
  &= \frac{N^2}{(4\pi)^2}\int\limits dx dy \state{x}{y}\statex{x}{y} \ ,
\end{align}
in agreement with \eq{normaliz-one},
using $c_0^2=\frac{N}{4\pi}$  and
\begin{align}
  \sum_{k}(2k+1) (-1)^{2\a+k}  \left\{
\begin{array}{lll}
 l & \a & \a \\
 k & \a & \a
         \end{array}\right\} = N\d_{l,0} \ .
\end{align}

\paragraph{The Laplacian $\Box$.}

In this case, \eq{Propagator-formula} gives
\begin{align}
 \Box &= \int dx dy \state{x}{y}\statex{x}{y}
    \sum_{k,l}(2k+1) k(k+1) (-1)^{l+k+2\a} c_l^2 \left\{
\begin{array}{lll}
 l & \a & \a \\
 k & \a & \a
         \end{array}
     \right\} \frac{2l+1}{4\pi} P_l(\cos\vartheta) \ .
\label{Laplace-kernelsum}
\end{align}
The sum over $k$ vanishes identically for $l\geq 2$ (from the representation theory
origin of the 6j symbols),
and for $l=0$ and $l=1$ one finds using the sum formulas \eq{6J-sums-1}
for 6j symbols
\begin{align}
 \Box &= N (N^2-1)\frac{1}{8\pi} \int dx dy 
 \state{x}{y}\statex{x}{y}
    \Big(c_0^2 -c_1^2 P_1(\cos\vartheta)\Big)
   \nn\\
   &= N^2 (N^2-1)\frac 12 \frac{1}{(4\pi)^2} \int dx dy 
   \state{x}{y}\statex{x}{y}
    \Big(1 - \frac{N + 1}{N-1} \cos\vartheta \Big) \nn\\
 &=  N^2 (N+1)^2\frac 14 \frac{1}{(4\pi)^2} \int dx dy 
  \state{x}{y}\statex{x}{y}
  \Big(|x-y|^2 \ -\frac{4}{N+1} \one\Big) \ ,
\end{align}
using \eq{c-i-explicit}. 
The second line can also be obtained directly from
\begin{align}
 \Box &= R_N^2 (X_L - X_R)^2 = 2R_N^2 (1 -  X_L X_R)\ ,
\end{align}
using  \eq{X-rep-coherent} for $X_L X_R$.
It is remarkable to find such a simple representation for the fuzzy Laplacian, since the classical Laplacian 
does not have a regular integral kernel.

\subsubsection{The string (or crossed) symbol of operators}
\label{sec:string-symbol-ops}

Similarly as the string kernel, the string symbol 
$\statex{x}{y} \cO \state{x}{y}$ \eq{string-symbol}
 can also be obtained  from Lemma \ref{lem:cross}.
 For functions of the Laplacian, this gives
\begin{align}
f(\Box) &= 
 \sum\limits_{l,m}f(l(l+1))\hat Y^l_m \otimes \hat Y^{l\dagger}_{m}  
 = \sum\limits_{l,k} f(l(l+1)) A^{lk}  \cC\big(\sum_n\hat Y^k_n \otimes \hat Y^{k\dagger}_{n} \big) \ ,
\end{align}
and therefore
\begin{align}
\statex{x}{y} f(\Box) \state{x}{y}
 &= \sum\limits_{l,k} f(l(l+1)) A^{lk} 
 \statex{x}{y}\cC\big(\sum_n\hat Y^k_n \otimes \hat Y^{k\dagger}_{n} \big)
   \state{x}{y}\nn\\
    &= \sum\limits_{l,k} f(l(l+1)) A^{lk} 
  \frac 1{c_k^2} Y^k_n(x) Y^{k*}_{n}(y)  \nn\\
    &= \sum\limits_{l,k} f(l(l+1)) A^{lk} \frac 1{c_k^2}
   \frac{2k+1}{4\pi} P_k(\cos\vartheta) \ .
 \end{align}
This is very similar to the string kernel \eq{Propagator-formula}, but the large $k$ modes are now suppressed rather than enhanced,
due to $\frac 1{c_k^2}\to 0$. We can therefore
use the approximation in \eq{Alk-def} for $A^{lk}$
for any $l$, so that 
\begin{align}   
 \statex{x}{y} f(\Box) \state{x}{y}   &\approx \frac{1}{2\a}\sum_{k,l}(2l+1) f(l(l+1))
   P_k\big(1-\frac{l^2}{2\a^2}\big)\frac 1{c_k^2}
    \frac{2k+1}{4\pi} P_k(\cos\vartheta) \ .
    \label{f-Box-expect}
\end{align}

Let us first compute the sum over $k$. 
As mentioned before, the coefficient $1/c_k^2$ effectively cuts off this sum at $k\sim\sqrt N$ due to \eq{c-n-asymptotic}.
Using the fact that $P_l(\cos\vartheta)\sim Y_0^l(\vartheta)$, the fact that ${\sum_k (2k+1)P_k(x)P_k(y)=2\d(x-y)}$ and the property of the symbol (\ref{symbol-Y}) we see that the sum over $k$ is equal to the symbol of the fuzzy $\d$-function (\ref{inner-coh}). This leads to  
\begin{align}   
 \statex{x}{y} f(\Box) \state{x}{y}
 &=\frac{1}{2\alpha}\sum_l (2l+1)f(l(l+1))\frac{1}{N}2 \d\big(1-\frac{l^2}{2\alpha^2}-\cos\vartheta \big)\approx\nn\\
 &\approx \frac{1}{\a}N \int_0^1 dy (2Ny+1) f\big(Ny(Ny+1)\big)\frac{1}{N}\delta\big(1-\frac{(Ny)^2}{2\a^2}-\cos\vartheta\big)=\nn
 \\&=\frac{1}{\a}\int du \frac{2\a\sqrt{2u}+1}{\frac{N}{\a}\sqrt{2u}} f\big(\a\sqrt{2u}(\a\sqrt{2u}+1)\big)\delta\big(1-u-\cos\vartheta\big)=\nn
 \\&=\frac{2\a}{N}\Big(1+\frac{1}{2\a|x-y|}\Big)f\big(\a|x-y|(\a|x-y|+1)\big)\ ,
 \label{string-symbol-general-k}
\end{align}
where we have used the normalization \eq{delta-normalization} and ${|x-y|^2=2(1-\cos\vartheta)}$. Unless ${x=y}$ the second term in the bracket does not contribute for large $N$ and we obtain
\begin{align}   
 \left(^x_{y} \right|f(\Box) \left|^x_{y} \right)   &\approx f\big(\a|x-y|(\a|x-y|+1)\big)\ .
\end{align}
 In particular for the propagator with $f(l(l+1)) = \frac{1}{l(l+1) + m^2}$, this gives
\begin{align}   
\boxed{
 \statex{x}{y}\frac{1}{\Box+m^2} \state{x}{y} =:\mathcal O_P(\cos\vartheta) \approx 
\frac{1}{\a^2|x-y|^2+m^2}\ .
}\label{string-symbol-large-N}
\end{align}
We have introduced a notation $\mathcal O_P(\cos\vartheta)$ for the (unapproximated) symbol of the propagator\footnote{Notice that we have obtained a similar expression to (\ref{PP-strings}), but without the $2L_{\rm NC}^2$ term. At this point it is not clear whether this is a shortcoming of our approximation or if expecting  $(E_{xy}+m^2)^{-1}$ would be too naive. We leave this question for future work.}. The above formula would however be at odds with \eq{delta-prop-1} in the case ${x=y}$. In this case the above derivation does not hold, and to check the consistency of \eq{f-Box-expect} and \eq{NC-Greens-1} we need to proceed differently.

A different formula for the propagator is obtained from \eq{f-Box-expect} by evaluating the $l$ sum first,
\begin{align}
 \sum_{l=0}^{2\a}  (2l+1) f(l(l+1))
   P_k(1-\frac{l^2}{2\a^2})
  \  &\approx \  2\a^2 \int_{-1}^{1} du f(2\a^2(1-u)) P_k(u)
\end{align}
for $u=1-\frac{l^2}{2\a^2}$.
For the propagator this yields
\begin{align}
 \sum_{l=0}^{2\a} \frac{2l+1}{l(l+1) + m^2}
   P_k(1-\frac{l^2}{2\a^2})
      \ &\approx \ \int_{-1}^{1} du \frac{1}{1-u}  P_k(u) 
\end{align}
as long as $m^2 \ll \a^2$.
This is divergent at $u=1$, but
as in \cite{Chu:2001xi}, we can evaluate the difference 
\begin{align}
 \int_{-1}^{1} du \frac{1}{1-u}  (P_k(u) - 1) &= -2 h(k),\qquad
  h(k) = \sum_{n=1}^k \frac 1n \approx \ln \big(k+ 1\big) + \gamma
\end{align}
where $\gamma$ is the Euler-Mascheroni constant.
Hence we can approximate 
\begin{align}
 \sum_{l=0}^{2\a} \frac{2l+1}{l(l+1) + m^2}
   P_k(1-\frac{l^2}{2\a^2})
  \  = 2\ln\Big(\frac{2\a}{k+1}\Big) \ ,
\end{align}
which is good for all $k$ including $k=0$.
Therefore
\begin{align}
 \statex{x}{y} \frac 1{\Box+m^2} \state{x}{y}
 \ \approx\ \frac{1}{\a}\sum\limits_{k=0}^{2\a}
 \ln\Big(\frac{2\a}{k+1}\Big)\frac 1{|c_k|^2}
    \frac{2k+1}{4\pi} P_k(\cos\vartheta)
    \label{string-symbol-general}
\end{align}
as long as $m^2 \ll \a^2$. which is an alternative formula to \eq{string-symbol-general-k}. This sum is well-behaved, because $c_k^{-2}$ provides an exponential cutoff at  $k \sim\sqrt{N}$. For coincident points $x=y$, we obtain
\begin{align}
 \statex{x}{x} \frac 1{\Box+m^2} \state{x}{x}
 &\approx \frac{1}{\a}\sum_{k}
 \ln(\frac{2\a}{k+1}) \frac{2k+1}{N} P_k(1)e^{- k^2/N} \nn\\
 &= \frac{1}{\a}\sum_{k}
 \ln(\frac{2\a}{k+1}) \frac{2k+1}{N} e^{- k^2/N} \nn\\
 &\approx - 4 \int_0^1 dx x\ln(x) e^{-N x^2} 
 = -  \int_0^1 du\ln(u) e^{-N u} \nn\\
 &\approx \frac 1N\ln(N)\ ,
 \label{symbol-propagator-local-full}
\end{align}
in agreement with \eq{delta-prop-1}.

These symbols are  reminiscent of 2-point functions in field theory.
However here, they are regular matrix elements of operators without singularities.
More generally, the distributional $n$-point functions of classical field theory are replaced
by regular matrix elements of operators in the noncommutative framework,
 extended by a non-local string sector.

\subsubsection{Trace computations}

The above string symbol allows to compute traces of the form
\begin{align}
 \Tr f(\Box) = \frac{N^2}{(4\pi)^2} \int dx dy \statex{x}{y} f(\Box) \state{x}{y}  \ .
 \label{Trace-string-full}
\end{align}
In particular,
\begin{align}
 \Tr \frac 1{\Box+m^2}  &= \frac{N^2}{4\pi} \int dy  \frac{1}{\a}\sum_{k}
 \ln\big(\frac{2\a}{k+1}\big)\frac 1{|c_k|^2}
    \frac{2k+1}{4\pi} P_k(\cos\vartheta) \nn\\
    &= \frac{N^2}{2\a} \sum_{k}
 \ln\big(\frac{2\a}{k+1}\big)\frac 1{|c_k|^2}
    \frac{2k+1}{4\pi}\int_{-1}^1 du  P_k(u) \nn\\
     &= \frac{N^2}{\a} \ln(2\a)\frac 1{|c_0|^2}
    \frac{1}{4\pi}  \nn\\
     &= 2 \ln(N)\ ,
\end{align}
noting that the double integral provides an explicit factor $4\pi$.
This correctly reproduces the  classical trace
\begin{align}
\Tr\frac 1{\Box+m^2} &= \sum_{l=0}^{N-1} \frac{2l+1}{l(l+1) + m^2} 
  \sim \int_0^{N} du \frac{2u}{u^2} = 2\ln(N) \ .
\end{align}

\paragraph{IR traces.}

Consider the trace over the IR modes up to $\L_\NC=1/L_\NC$
\begin{align}
\Tr_{\L_\NC}\frac 1{\Box+m^2} &= \sum_{l=0}^{\sqrt{N}} \frac{2l+1}{l(l+1) + m^2} 
  \sim \int_0^{\sqrt{N}} du \frac{2u}{u^2} = \ln(N) \ .
\end{align}
This is reproduced by the local string modes,
\begin{align}
 \Tr_{\L_\NC}\frac 1{\Box+m^2} &\sim \frac{N}{4\pi}
 \int dx \statex{x}{x} \frac 1{\Box+m^2} \state{x}{x} 
  = N \statex{0}{0} \frac 1{\Box+m^2} \state{0}{0}
  \sim \ln(N)
\end{align}
using \eq{symbol-propagator-local-full}.
This can be viewed as trace over the short or local string modes 
in \eq{Trace-string-full},
which thus reduces to the classical trace with cutoff $\L_{NC}$.
Indeed as discussed in section \ref{sec:kinematics-strings}, 
the modes $\state{x}{y}$ 
with $x\approx y$
 are in one-to-one correspondence with functions 
with UV cutoff $\L_{NC}$. These is useful for 
 computations of the induced gravity action in matrix models \cite{Steinacker:2021yxt}.

\subsection{Off-diagonal string symbol and semi-classical 
   representation of operators}\label{sec:string-IR-op}

In analogy to section \ref{sec:string-IR reg-quant}, 
we obtain a well-behaved representation of the propagator 
in terms of  off-diagonal string modes as follows
\begin{align}
  (\Box + m^2)^{-1}  = \int dx\,dy\,dx'\,dy'\, \state{x}{y} K(xy;x'y') \statex{x'}{y'}\ ,
  \label{off-diag-string-prop}
\end{align}
with 
\begin{align}
 K(xy;x'y') := \frac{N^4}{(4\pi)^4} 
 \statex{x}{y} (\Box + m^2)^{-1} \state{x'}{y'} \ .\label{K:general}
\end{align}
This follows immediately from the completeness relation 
\begin{align}
 \one &= \frac{N^2}{(4\pi)^2}\int\limits dx dy \state{x}{y} \statex{x}{y} \ ,
 \label{completeness-End}
\end{align}
which follows from \eq{normaliz-one}, or from $SO(3)_L \times SO(3)_R$ invariance. 
In contrast to the diagonal form \eq{quant-End-End},
the function $K(xy;x'y')$ does not exhibit any 
oscillatory behavior, and is strongly peaked in the two regimes:
\begin{align}
\statex{x}{y} (\Box + m^2)^{-1} \state{x'}{y'}
 \ &\approx \  \frac{1}{\frac{N^2}{4}(x-y)^2 + m^2 }\braket{x}{x'}\braket{y'}{y}   \nn\\
 &\quad  +  G_\L(x-x') \braket{x}{y}\braket{y'}{x'} \ .
 \label{string-prop-two-regimes}
\end{align}

Here $G_\Lambda(x-x')$ is the UV-regulated classical propagator \eq{prop-class-approx} in position space.
Hence $K(xy;x'y')$ is  ''almost-diagonal`` in the s-channel and t-channel,
with a different localization behavior in the IR and UV modes: 
in the IR regime we should recover the
classical propagator, and in the UV regime  the string propagator. 
This follows from the considerations 
in sections \ref{sec:kinematics-strings}.

\begin{figure}
	\begin{center}
	\includegraphics[width=0.35\columnwidth]{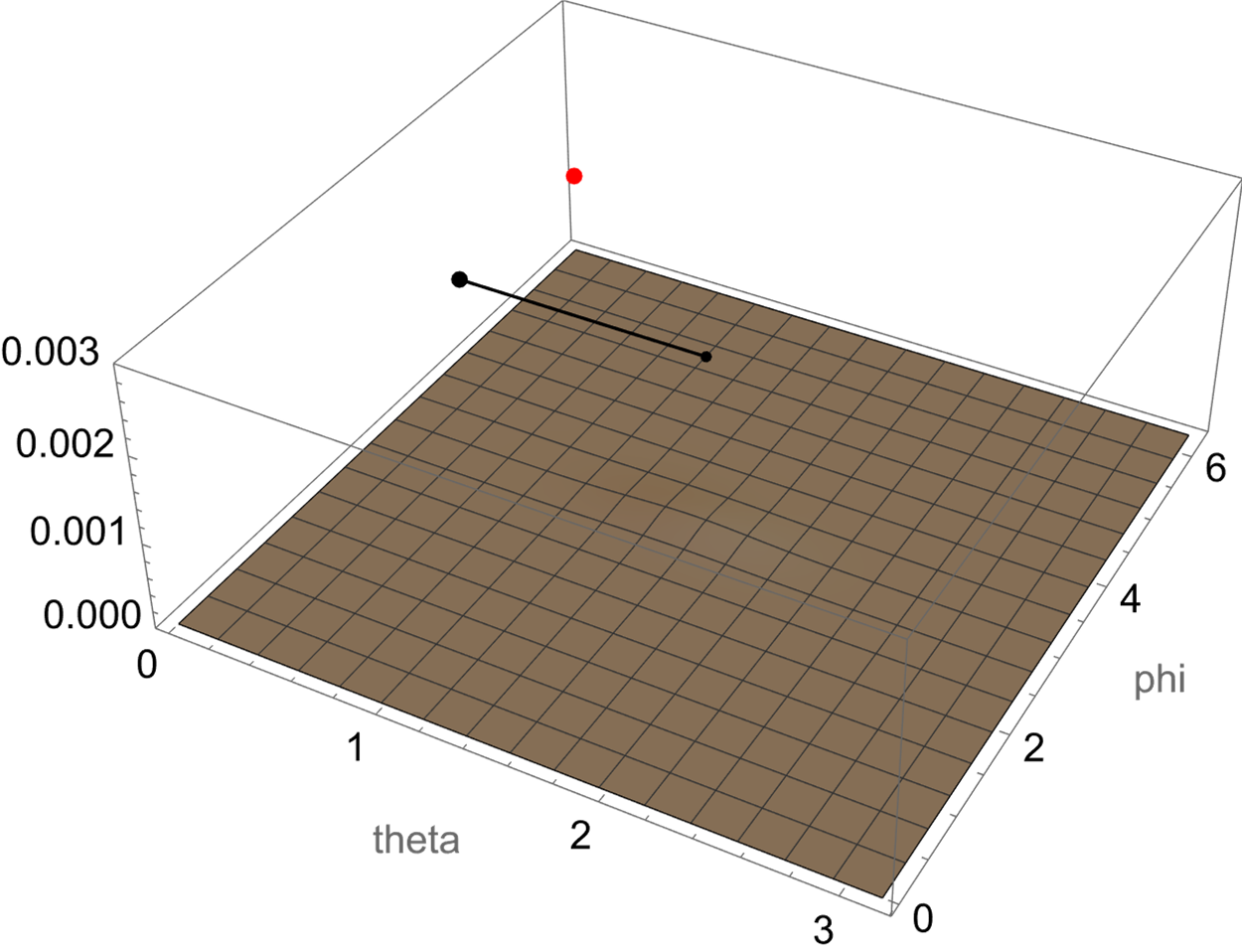}\hskip 35pt
	\includegraphics[width=0.35\columnwidth]{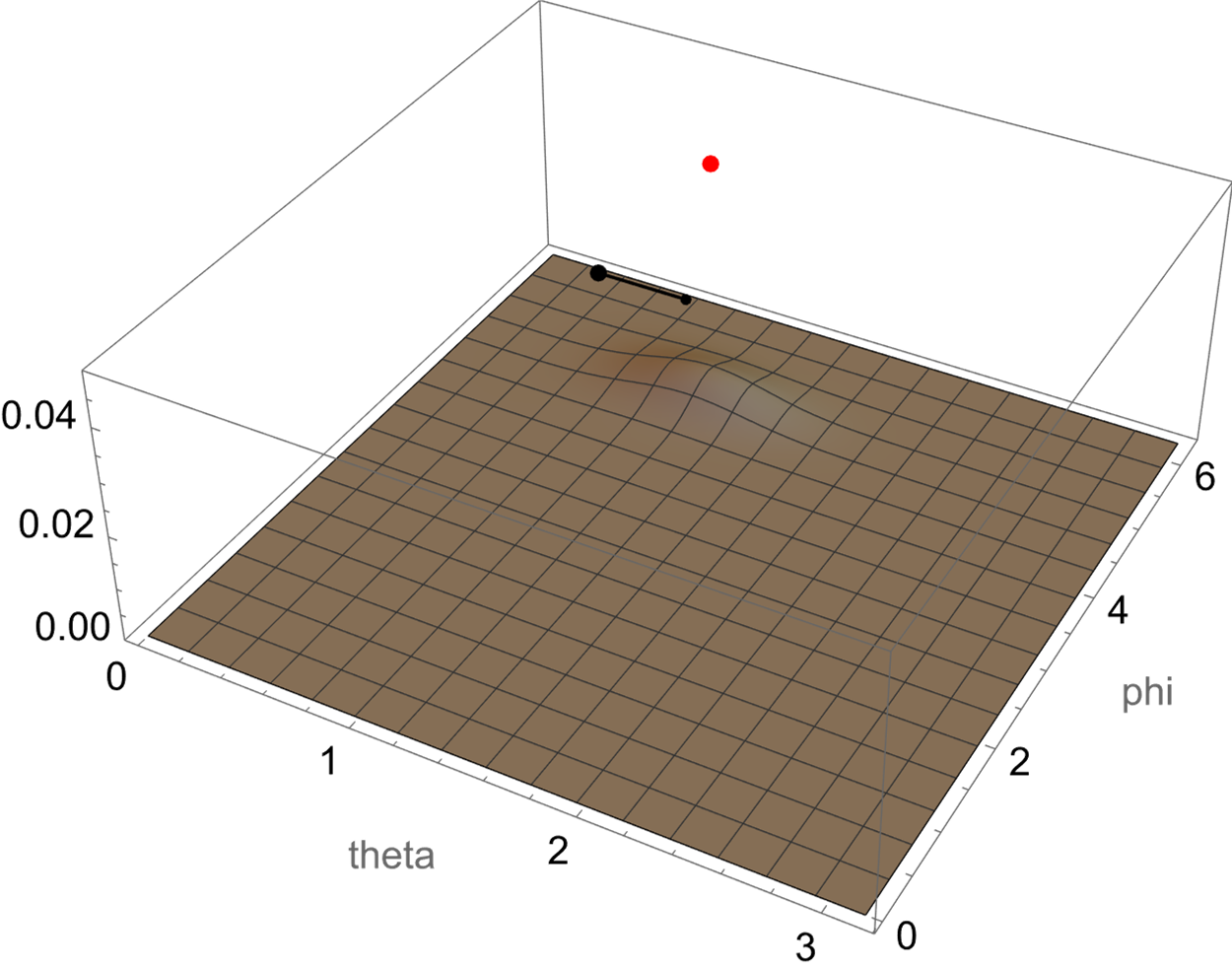}\\
	\includegraphics[width=0.35\columnwidth]{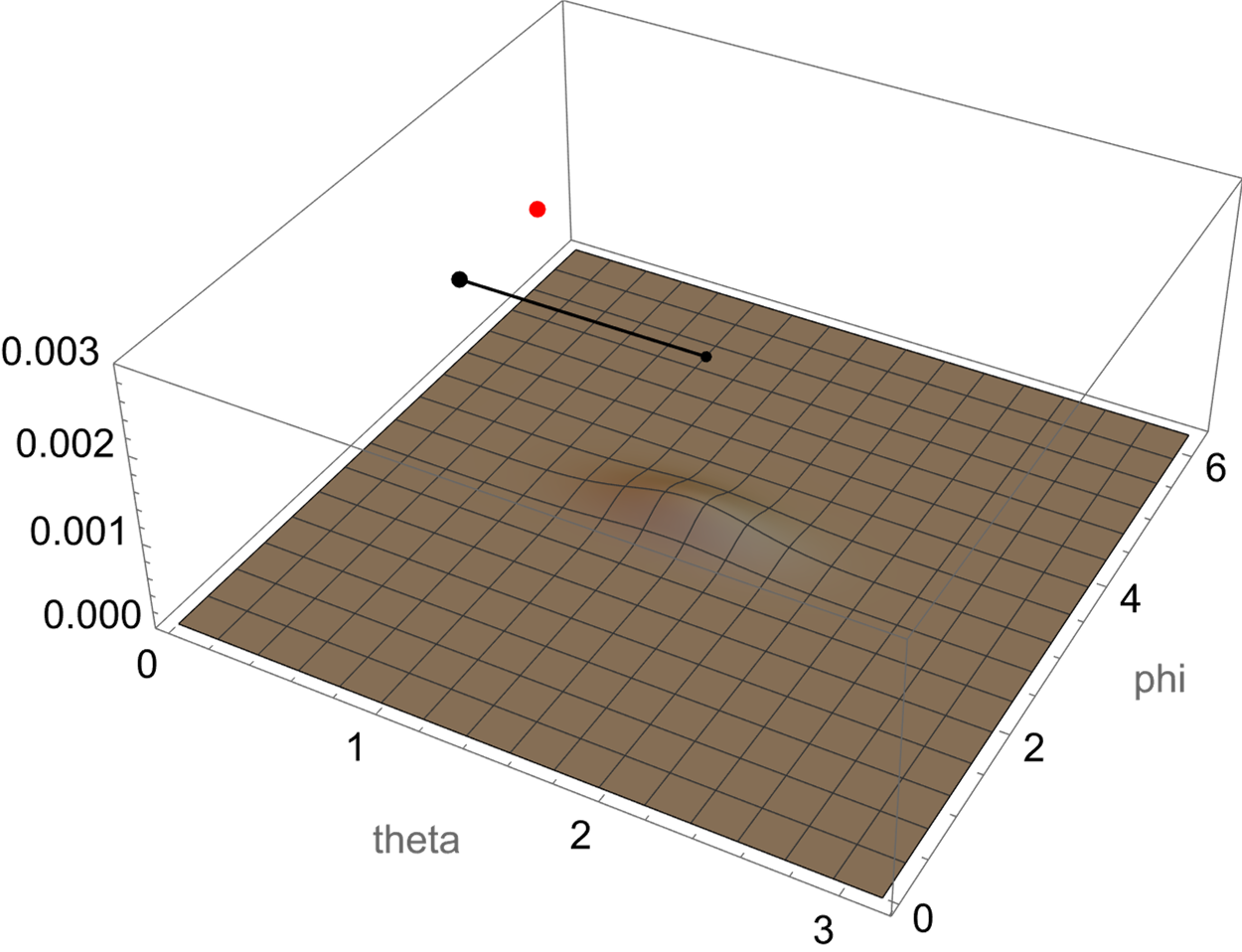}\hskip 35pt
	\includegraphics[width=0.35\columnwidth]{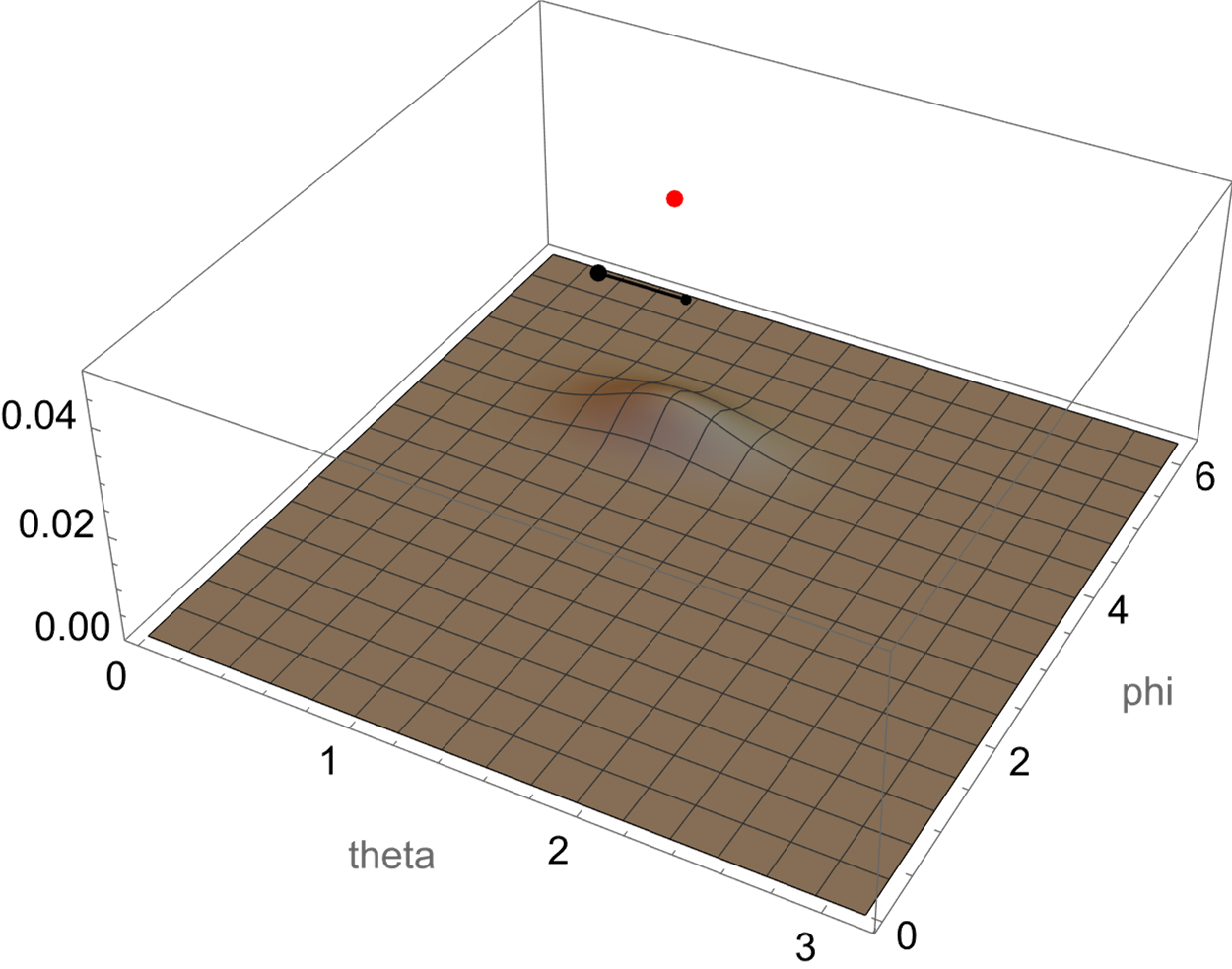}\\
	\includegraphics[width=0.35\columnwidth]{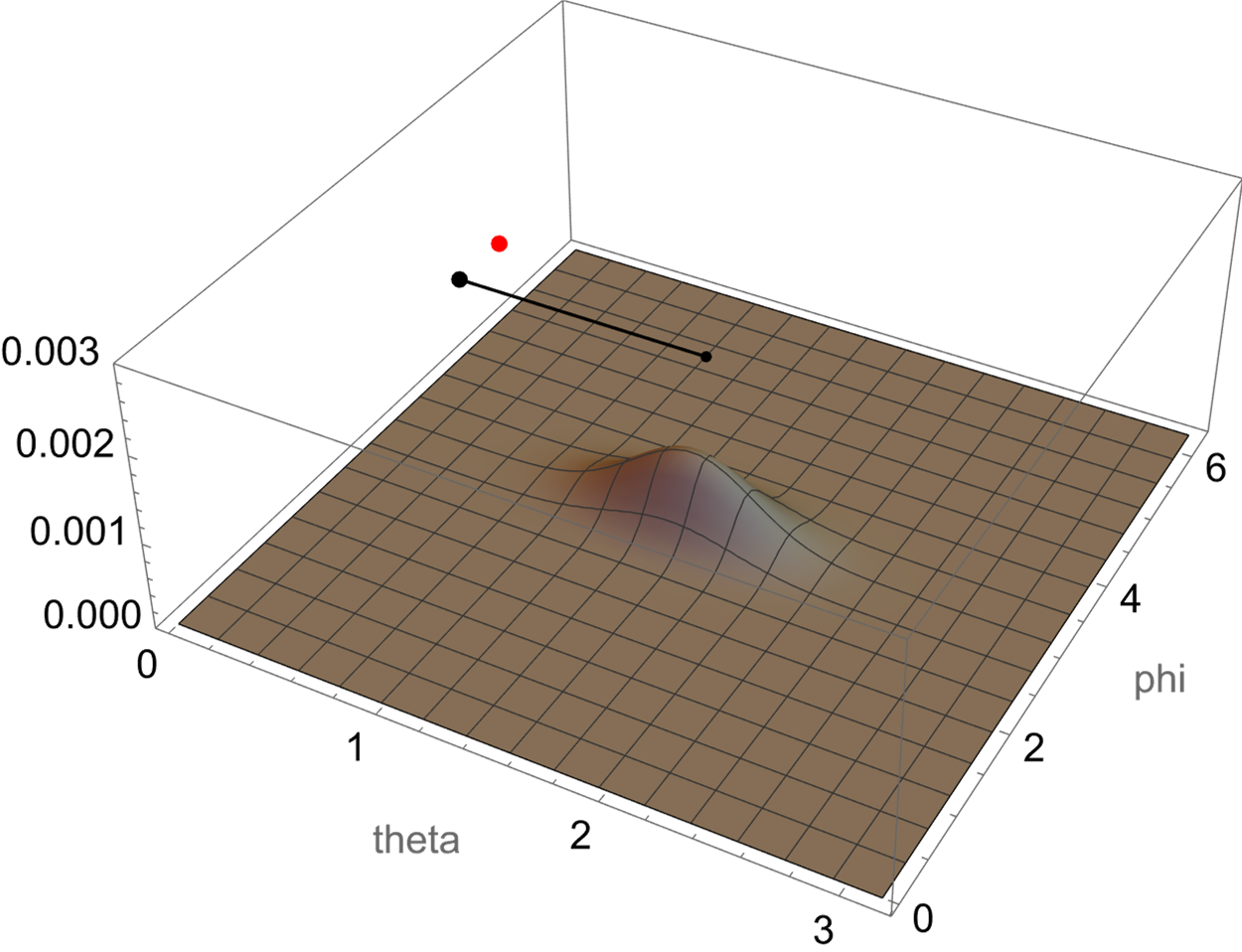}\hskip 35pt
	\includegraphics[width=0.35\columnwidth]{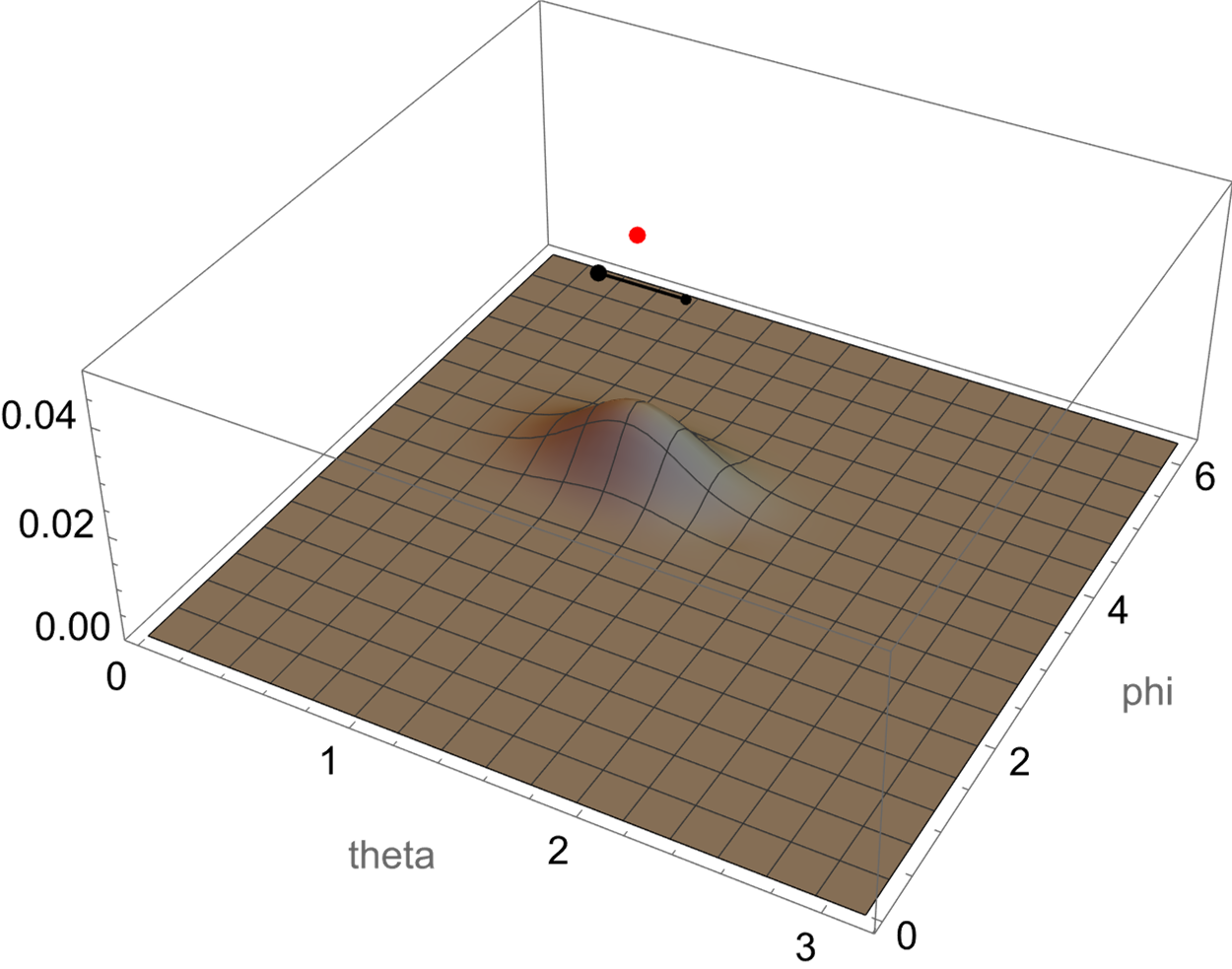}\\
	\includegraphics[width=0.35\columnwidth]{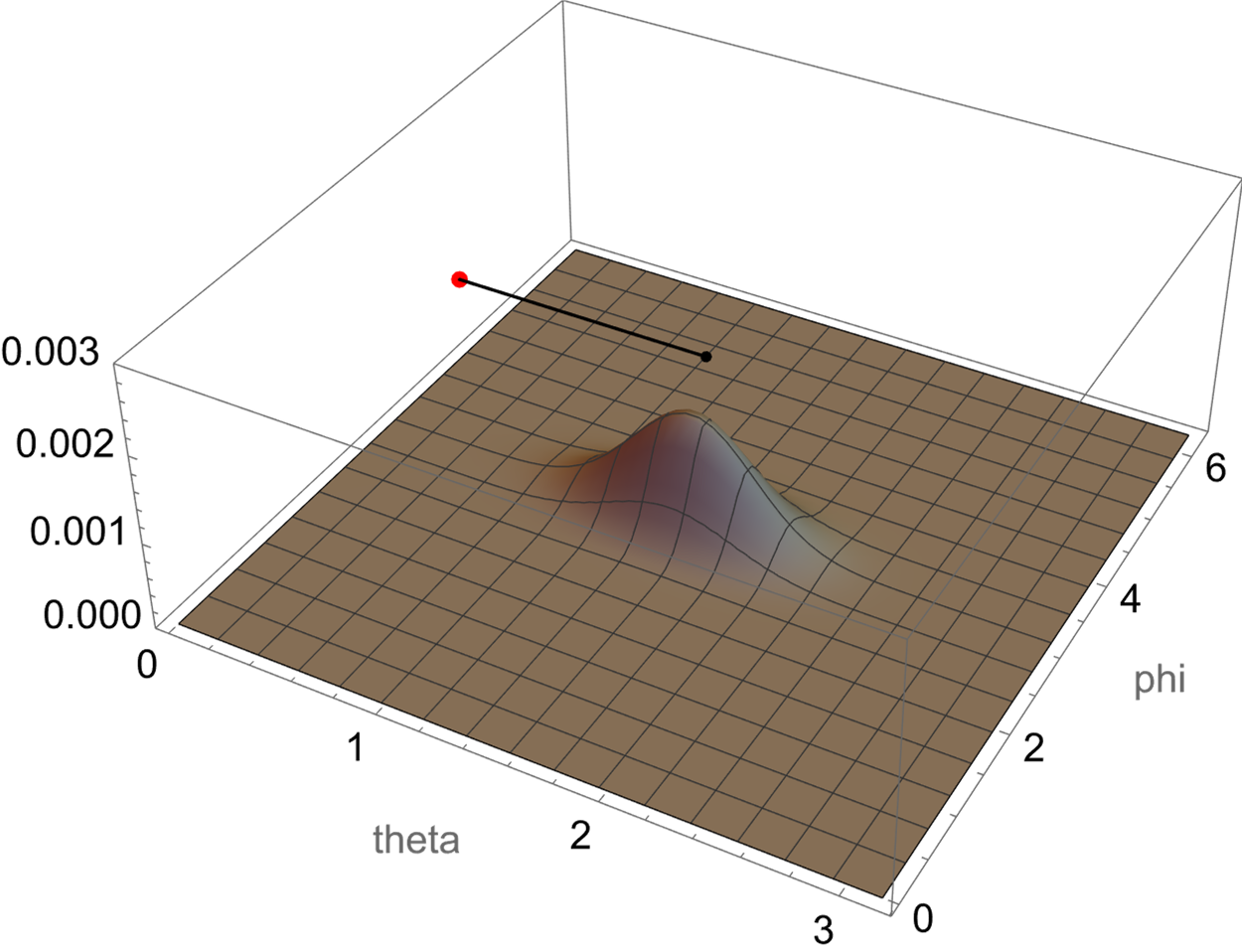}\hskip 35pt
	\includegraphics[width=0.35\columnwidth]{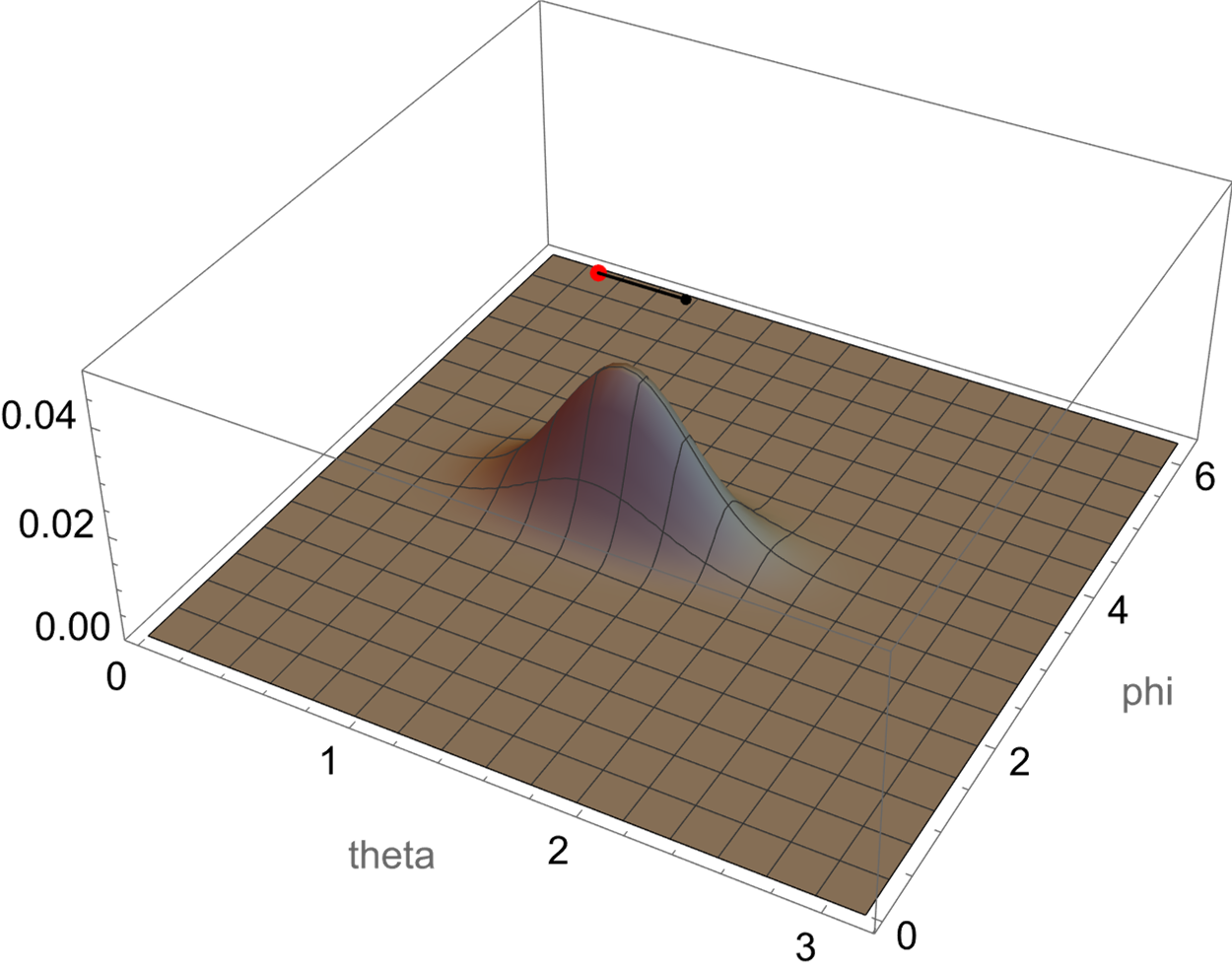}\\
	\end{center}
	\caption{Two sets of plots for a general off-diagonal symbol (\ref{K:general}) for the ${N=52}$ case. The two black connected dots denote endpoints of the string state $\state{x}{y}$, the upper label $x$ by the larger dot. The position of the red dot denotes the position of the label $x'$ of the string state $\state{x'}{y'}$. The two variables are all possible locations of the final end point $y'$ on the sphere and the graph represents the value of $K(x,y;x',y')$ for the fixed $x,y,x'$ and varying $y'$.
	The angular separation of the points in the first column of plots is $a$, in the second $b$, the sphere has unit radius. Since in this case $L_{NC}\approx 0.2$, the first case is in the UV-regime, the second in the IR regime.
	We can clearly see that in the IR regime, the symbol is localized around the ${x=x'=y'}$, while in the UV regime the localization is around the ${x=x',y=y'}$ case.
	}
	\label{Kplots}
\end{figure}

It remains to show that  $K(xy;x'y')$
is not oscillating and has the desired properties.
We already computed the diagonal VEV for ${x=x'}$ and ${y=y'}$, consistent
with the above picture.
Moreover, we can compute the string expectation value numerically
for general positions $x,y,x',y'$, see figure \ref{Kplots}.
It indeed displays the expected strong decay for non-coincident points 
for separations larger than $L_{NC}$.
In particular, we see that 
\begin{align}
 \statex{x}{y} (\Box + m^2)^{-1} \state{x'}{y'}
\end{align}
is peaked for ${x\approx y}$ and ${x'\approx y'}$ if $(x,y)$ 
are very far from $(x',y')$; this is the expected 
behavior of the classical propagator.
There is an interesting crossover-behavior between the two regimes
when $x,y,x'$ from an equilateral triangle. Then two peaks arise, 
one for $y'\approx x'$ and 
one for $y'\approx y$. Overall, the numerical results support the above picture.

The smoothness of $K(xy;x'y')$ can be justified
by an argument similar as in section \ref{sec:string-IR reg-quant}.
First, the Hilbert-Schmidt norm of the propagator 
is easily computed as
\begin{align}
 \|(\Box + m^2)^{-1}\|_{\rm HS} = \sum_l  \frac {2l+1}{l(l+1) + m^2}
 \approx 2\ln(N)\ ,
\end{align}
as long as $m$ is sufficiently small. 
Furthermore, the derivatives of the string modes are bounded by \eq{del-coherent-bound}
\begin{align}
  \|\del_\mu\left|^{x}_{y} \right) \|^2 
  \leq \frac{1}{L_\NC^2} \ .
\end{align}
Then an argument along the lines of \eq{del-strigsymb-est} gives 
the following bound for the derivative of string symbol of the propagator
 \begin{align}
  \Big|\frac{\del}{\del y^\mu} \left(^{x}_{y} \right| (\Box + m^2)^{-1}
  \big|^{x'}_{y'} \big) \Big|
  &= \big| \left(^{x}_{y} \right|(\Box + m^2)^{-1}\del_\mu\left|^{x}_{y} \right) \big| \nn\\
  &\leq  \|(\Box + m^2)^{-1} 
   \left|^{x}_{y} \right) \|
  \|\del_\mu \left|^{x}_{y} \right) \| \nn\\
  &\leq  \|(\Box + m^2)^{-1}\|_{\rm HS} \frac{1}{L_\NC} \nn\\
  &\leq  \frac{ 2\ln(N)}{L_\NC} \ .
 \label{del-prop-est}
 \end{align}
 Therefore the off-diagonal representation of the propagator
 is smooth on scales below $L_\NC$, and at longer scales it is clearly 
 dominated by the two diagonal regimes \eq{string-prop-two-regimes}.
 Note that the $\ln(N)$ factor in \eq{delta-prop-1} is now properly recovered.
 
This representation will be very useful to derive 
quantum effective actions in NCFT.

\section{Loop computations and (non)locality in NCFT}
\label{sec:loops}

After the above discussion and construction of the string representation of the operators on the fuzzy sphere, most importantly the propagator, we are in a prime position to discuss the formulation of the scalar field theory on the fuzzy sphere. This is the simplest non-trivial setting, in which these notions apply and yield some  important results.

The real scalar field theory on the fuzzy sphere is given by a suitable adaption of the standard commutative action for the scalar field
\begin{align}
S=\int dx\,\left( \frac{1}{2}\phi(x)\Box \phi(x)+\frac{1}{2}m^2\phi(x)^2+\frac{g}{4!}\phi(x)^4\right)\ ,
\label{action-class}
\end{align}
where we have chosen the quartic interaction. In the
fuzzy case, the real function $\phi$ is replaced by a hermitian matrix $\Phi\in\End(\cH)$, the integral is an appropriately rescaled trace, and the Laplacian is the double commutator with the three $SU(2)$ generators defining the sphere (\ref{Js}). Then  the fuzzy action is given by
\begin{align}\label{fuzzy-action}
S=\frac{4\pi}{N}\tr\left( \frac{1}{2}\Phi[J^a_{(N)},[J_{a(N)},\Phi]]+\frac{1}{2}\mu^2\Phi^2+\frac{g}{4!}\Phi^4\right)\ .
\end{align}
The quantization of this model is defined in terms of a matrix "path" integral over 
the space of hermitian matrices 
\begin{align}
\Phi = (\phi^i_j)  \quad \in \cA:= \End(\cH) \ ,
\end{align}
denoting the algebra of functions on the fuzzy sphere with $\cA$ for better readability.
Then
the correlators or $n$-point functions are obtained as usual from a generating function
\begin{align}
 Z[\cJ] = \int_\cA D\Phi\, e^{-S[\Phi] + \tr \Phi \cJ} = e^{-W[\cJ]} \ .
\end{align}
The perturbative 
expansion of a correlator is given by the sum of contractions of the interaction vertices with the propagator, which is viewed more abstractly as an  element in $\cA \otimes \cA^* \cong End(\cA)$:
\begin{align}
 \langle \phi^i_j\, {\phi^*}^k_l \rangle_0 \quad \in \cA \otimes \cA^*
 \label{prop-abstract}
\end{align}
represented by a double line  
starting at $({}^k_l)$ and ending at $({}^i_j)$. 
The quartic product in the interaction is however no longer invariant under arbitrary permutations  of the external legs, but only under cyclic permutations.
It is well-known that this leads to ribbon (Feynman) diagrams, with propagators represented by double lines. The distinction between  planar and nonplanar diagrams and their phase factors then leads to the well known phenomenon of the UV/IR mixing \cite{Minwalla:1999px}.

The conventional way to evaluate these diagrams
is to express the field in terms of the basis $\hat Y^l_m$ (\ref{hat-Y-CG-1}), which diagonalizes the kinetic term.
The explicit computations using this approach however become very difficult beyond one-loop, since they involve complicated group theoretical factors. Here, we will evaluate these diagrams in a different way in terms of string modes and the string symbol of the propagator. 
This is achieved by observing that any ribbon diagram is defined in a basis-independent way as a canonical contraction of propagators \eq{prop-abstract} with  vertices, which can be viewed as  element in $\cA^{\otimes 4}$ (using the canonical identification $\cA\cong\cA^*$ given by the trace). 
Representing the propagator in terms of its 
off-diagonal\footnote{Recall that the off-diagonal string symbol is well-behaved and non-oscillatory, in contrast to the diagonal string representation \eq{Propagator-formula}.} string symbol
\eq{off-diag-string-prop}
leads to a ribbon diagram, where the propagator is represented by a double line labeled by position space variables
\begin{align}
	\adjincludegraphics[height=1.25cm,valign=c]{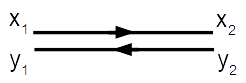}= \statex{x_2}{y_2}\frac{N}{4\pi}\left( \Box + m^2\right)^{-1}\state{x_1}{y_1}\approx \frac{N}{4\pi}\,\frac{1}{\frac{N^2}{4}|x_1-y_1|^2+m^2}
	\braket{x_2}{x_1}\braket{y_1}{y_2}\ ,\label{rules_propagator}
\end{align}
using the approximation (\ref{string-symbol-large-N}) in the last form. 
The vertex then takes an almost-local form in position space
\begin{align}
	\adjincludegraphics[height=3cm,valign=c]{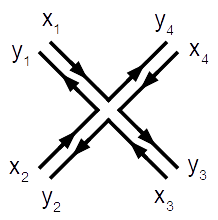}= \frac{4\pi}{N} \frac{g}{4!}
	\braket{y_1}{x_2}\braket{y_2}{x_3}\braket{y_3}{x_4}\braket{y_4}{x_1} \ .
\label{rules_vertex}
\end{align}
Let us point out that the factors of $4\pi/N$ in these two expression come from the normalization of the fuzzy action (\ref{fuzzy-action}). The diagrams are thus formed by joining the vertices with propagators and summing over the loop degrees of freedom in the same way as one does for the ribbon graphs of matrix models. The labels  $x$ etc. play the role of matrix indices $i\in\{1,\ldots,N\}$, 
and internal indices are to be integrated over position space $S^2$, replacing 
\begin{align}
\sum_i & \to \frac{N}{4\pi}\int dy\ . 
\label{line-factor}
\end{align}
The factors $\braket{y}{x}$ play the role of $\langle i|j\rangle$ and behaves similar to a delta-function or Kronecker delta, but including a phase. This phase will always cancel for internal labels which can be evaluated using 
the completeness relation, such as 
\begin{align}
 \frac{N}{4\pi}\int dy 	\braket{x}{y}\braket{y}{z}
  = \braket{x}{z}
  \label{complete-feynman}
 \end{align}
 which replaces $\sum_j	\braket{i}{j}\braket{j}{k}
  = \braket{i}{j}$ for the standard spin basis.

Another equivalent way to obtain this representation  is to insert the completeness relation \eq{completeness-End} \begin{align}
 \one &= \frac{N^2}{(4\pi)^2}\int\limits dx dy \left|^x_{y} \right) \left(^x_{y}\right| \qquad 
 \in    \cA \otimes \cA^*
\end{align}
at the end of each propagator.

This representation of amplitudes or diagrams is extremely useful, because all objects are well localized in position space, and the 
integrals are very easy to evaluate and to interpret.
We can simplify these rules further by observing that
the internal factors $\braket{y}{x}$ 
impose the conservation of position labels $x$ along the lines, and integrating them out using \eq{complete-feynman} leaves only the factors $\braket{y}{x}$ for the end-points of the lines, i.e. the external lines of the diagram.
For edges with no endpoints, we are left simply with ${\braket{x}{x}=1}$, and each closed internal line leaves a label to be integrated over. 
This will be illustrated  in \eq{planar-1-loop} below.
These steps can be subsumed in terms of the following {\bf reduced ribbon Feynman rules}:
\begin{enumerate}
\item the lines of the ribbon diagrams carry
position labels $x$, which are conserved along the arrows along each line. External lines with in-going label $x$ and out-going label $y$
    acquire a factor $\braket{y}{x}$
    \item each internal propagator  with line labels $x$ and $y$ acquires a factor\footnote{The notation for the string state symbol of the propagator $\mathcal O_P$ has been introduced in \eq{string-symbol-large-N}.}     $$\frac{N}{4\pi} \mathcal O_P(\cos\vartheta)
     \approx \frac{N}{4\pi}\frac{1}{\frac{N^2}{4}|x-y|^2+m^2}\ ,$$
    \item each  vertex  acquires a factor 
    $\frac{4\pi}{N} \frac{g}{4!}$,
    \item all internal closed lines are integrated over with measure $\frac{N}{4\pi}\int dx$ .
\end{enumerate}
These rules are based on the approximation
\eq{rules_propagator} for the propagator, which holds in the UV regime. Therefore they
correctly reproduce all diagrams which are  dominated by the UV regime, which is the typically the case in quantum field theories.
Diagrams which are dominated by the IR physics can be computed either using the standard rules of (noncommutative) QFT without encountering oscillatory behavior,
or by refining the approximation for the string kernel for the propagator in \eq{string-prop-two-regimes}, recalling \eq{wavepacket}.

Apart from the non-trivial dependence of the propagator on the difference $|x-y|$ of the labels of the  double line,
these rules are completely analogous to the well-known 't Hooft rules for matrix models, which are recovered in the limit $m^2 \to \infty$.
More precisely, setting
\begin{align}
    m^2=N \tilde m^2\
    \label{large-mass}
\end{align}
the action \eq{fuzzy-action} takes the form
\begin{align}\label{fuzzy-action-tHooft}
S= 4\pi\,\tr\left( \frac{1}{2N}\Phi[J^a_{(N)},[J_{a(N)},\Phi]]
+\frac{1}{2}\tilde m^2\Phi^2+\frac{g}{4! N}\Phi^4\right)\ .
\end{align}
This has the standard form of single-matrix models where $g$ plays the role of the 't Hooft coupling, with an extra kinetic term $\frac 1N[J,[J,.]]$. Without this term, the ribbon graphs are known to scale like $ g^{V} N^{2-2H}$ \cite{Brezin:1977sv}, where $V$ is the number of vertices and $H$ is the genus. Then planar diagrams dominate, and the theory simplifies. 
The kinetic term leads to an extra dependence on 
$|x-y|$ in the propagator \eq{rules_propagator}. More precisely, the propagator is suppressed for large separation $|x-y| = O(1)$, while the mass term $\tilde m$ dominates for the local contributions $|x-y|^2 = O(\frac 1N)$, where
the usual 't Hooft scaling is recovered. It would be interesting to explore the resulting behavior in more detail. Here, we restrict ourselves to the 
simple observation that the diagrams are bounded from above by the 't Hooft scaling  $g^{V} N^{2-2H}$ of pure matrix model.

It is also instructive to rewrite the action directly in the non-local string basis as follows
\begin{align}
S\,=\,&\frac{4\pi}{N}\left( \frac{N}{4\pi}\right)^4\int dx_1\,dy_1\,dx_2\,dy_2\,\phi(x_1,y_1)\phi(y_2,x_2)\statex{x_2}{y_2}\frac{1}{2}\left( \Box + m^2\right)
\state{x_1}{y_1}+\nonumber\\
&+\frac{4\pi}{N}\frac{g}{4!}\left( \frac{N}{4\pi}\right)^8\int dx_1\,dy_1\,dx_2\,dy_2\,dx_3\,dy_3\,dx_4\,dy_4\,\phi(x_1,y_1)\braket{y_1}{x_2}\phi(x_2,y_2)\nonumber\\&\ \ \ \ \braket{y_2}{x_3}\phi(x_3,y_3)\braket{y_3}{x_4}\phi(x_4,y_4)\braket{y_4}{x_1} \ .
\label{NCFT-bare-action} 
\end{align}
This is an action for a bi-local field $\phi(x,y) = \langle x|\phi|y\rangle$. Note that the nonlocality of the star product is explicitly transformed into nonlocality of the above action, even in the large $N$ limit.
However, the restriction of the modes  $\phi(x,y)$
to the IR regime 
with cutoff at $L_{NC}\approx 1/\sqrt N$
as derived in section \ref{sec:string-IR reg-quant}
must be imposed by hand.
This is easy to do for external fields, but 
more problematic in the path integral. Therefore
the Feynman rules are more easily derived using the previous approach.

\begin{figure}
	\begin{center}
	\includegraphics[width=0.3\columnwidth]{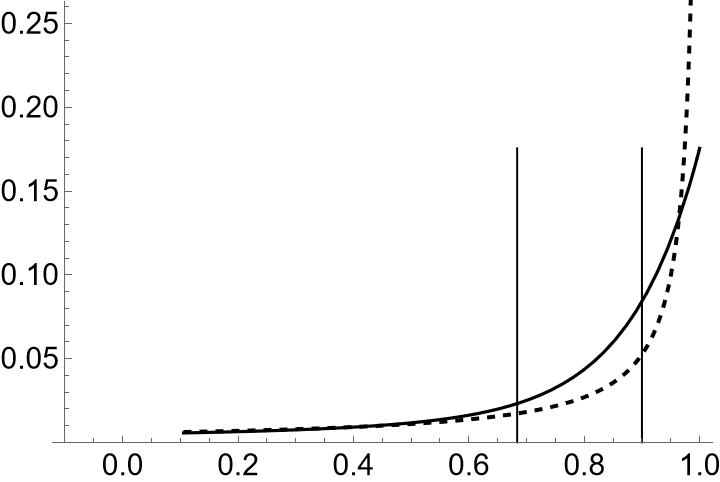}
	\includegraphics[width=0.3\columnwidth]{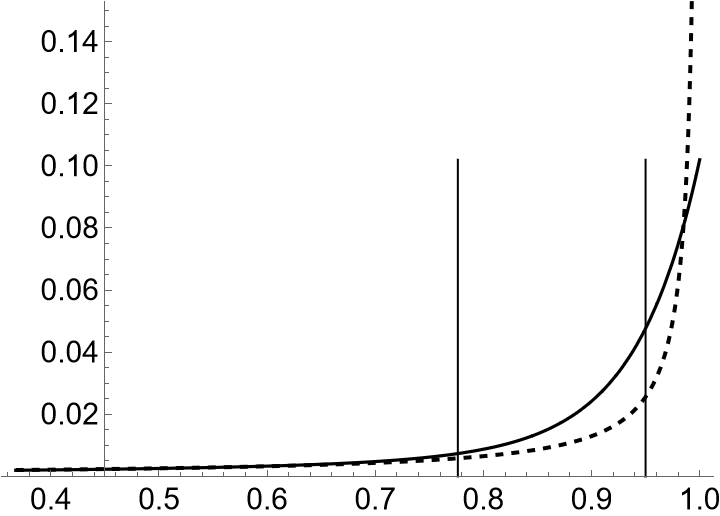}
	\includegraphics[width=0.3\columnwidth]{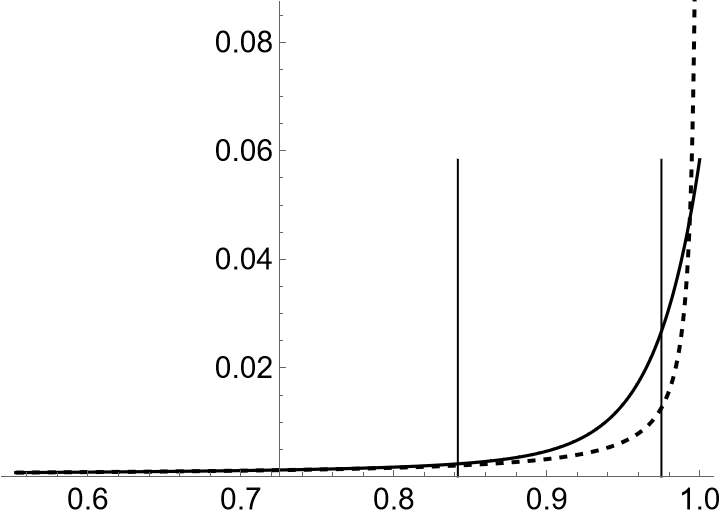}
	\includegraphics[width=0.3\columnwidth]{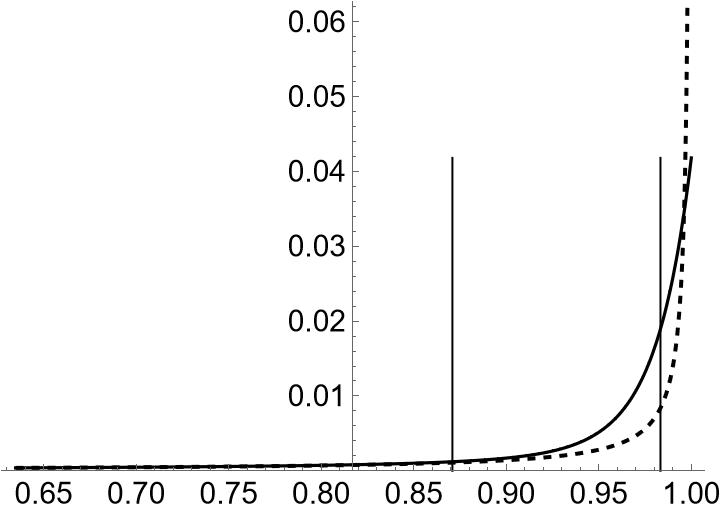}
	\includegraphics[width=0.3\columnwidth]{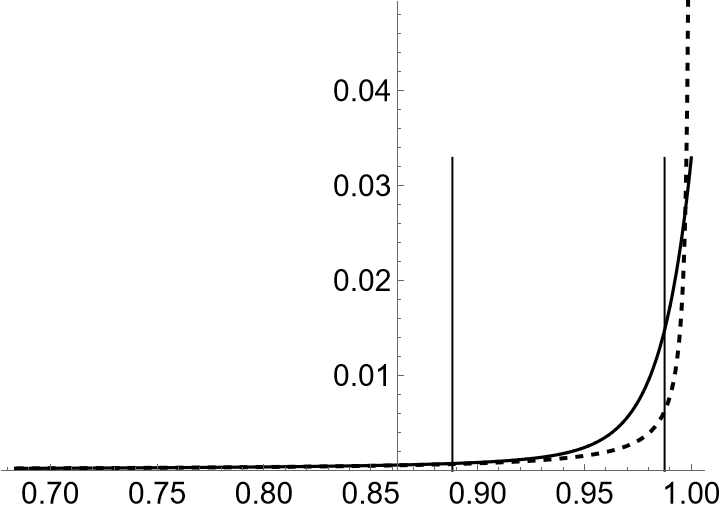}
	\includegraphics[width=0.3\columnwidth]{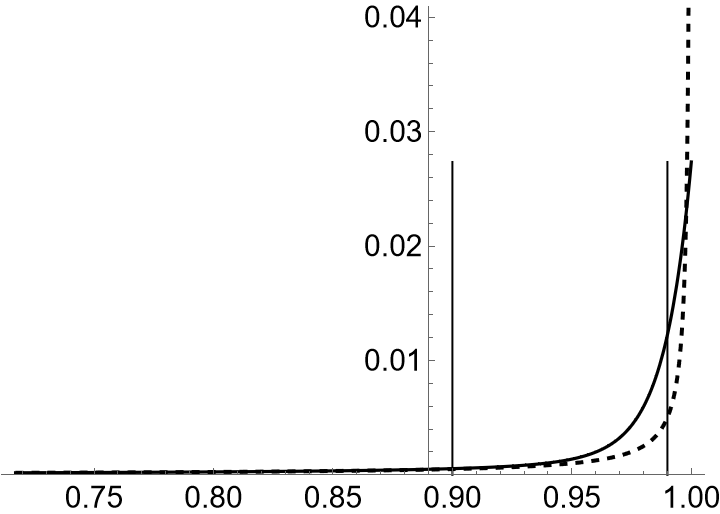}
	\end{center}
	\caption{Graphs of the exact formula \eq{f-Box-expect} - the black line, and the approximation \eq{string-symbol-large-N} - the dashed thick line, for ${N=20, 40, 80, 120, 160, 200}$ (left to right, top to bottom). The parameter on the horizontal axis is ${u=\cos\vartheta}$ and the two vertical lines are at ${u=1-\sqrt 2/\sqrt N}$ and ${u=1-2/N}$. We can see that the approximation is reliable only up to roughly  the first of these values. Note that the $u$ axis is shifted and rescaled such that the relevant region is visible.}
	\label{fig1}
\end{figure}

Let us comment on the validity of the approximation \eq{string-symbol-large-N}. It is strictly valid only after the large $N$ limit is taken. For finite $N$ there are going to be corrections and if these corrections are large, the approximation might not be sufficiently precise. We can see this by numerically evaluating the sum \eq{f-Box-expect} for the propagator for finite values of $N$ and comparing the result with the approximation \eq{string-symbol-large-N}. This is show in Figure \ref{fig1}. As we can see, the approximation is valid up to ${1-\cos\vartheta\gtrsim 1/\sqrt N}$ or ${|x-y|\gtrsim N^{-1/4}}$.
This is however not sufficient for our purposes, since we need to consider string states with lengths all the way to $1/\sqrt N \approx L_{NC}$. We will also explicitly see how this would cause an issue in section \ref{sec:one-loop-two-point}.

\begin{figure}
	\begin{center}
	\includegraphics[width=0.3\columnwidth]{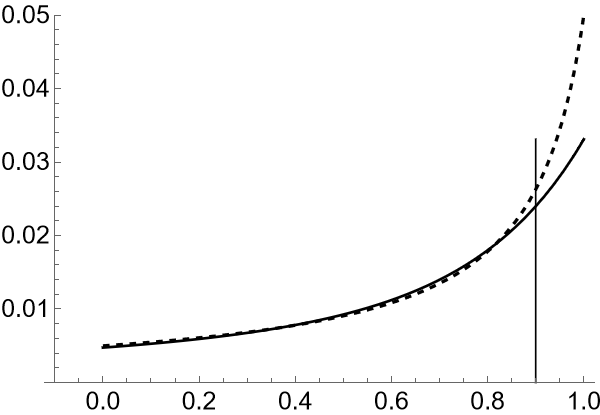}
	\includegraphics[width=0.3\columnwidth]{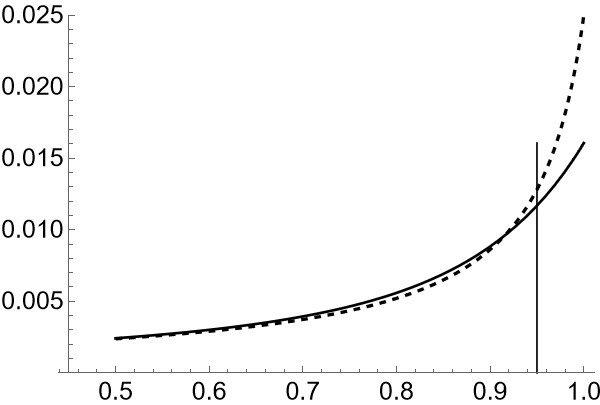}
	\includegraphics[width=0.3\columnwidth]{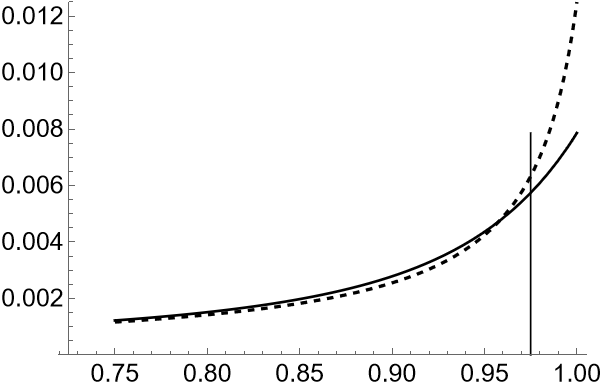}
	\includegraphics[width=0.3\columnwidth]{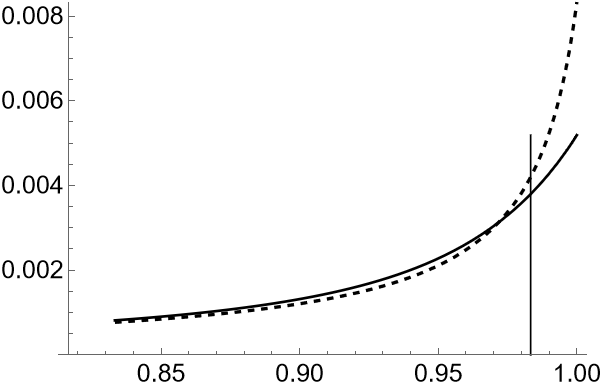}
	\includegraphics[width=0.3\columnwidth]{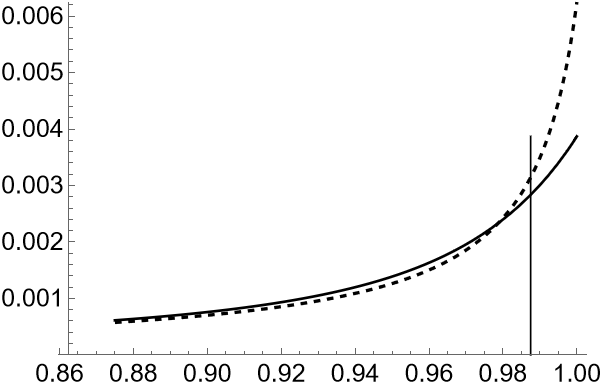}
	\includegraphics[width=0.3\columnwidth]{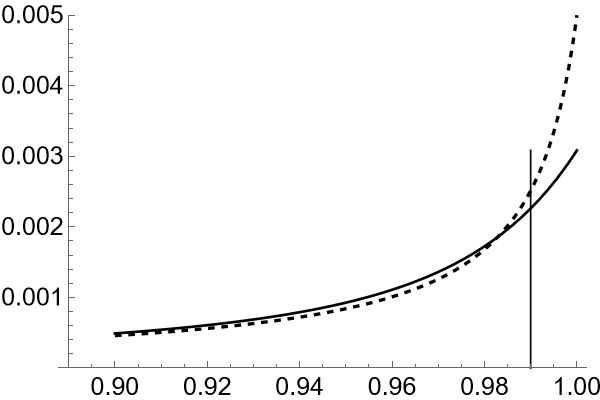}
	\end{center}
	\caption{Graphs of the exact formula \eq{f-Box-expect} - the black line, and the approximation \eq{string-symbol-large-N} - the dashed thick line, for ${N=20, 40, 80, 120, 160, 200}$ (left to right, top to bottom) in the large mass approximation \eq{large-mass} The parameter on the horizontal axis is ${u=\cos\vartheta}$ and The vertical line is at ${u=1-2/N}$. We can see that in the large mass approximation the approximation is reliable all the way to this value. As in the figure \ref{fig1} the $u$ axis is shifted and rescaled such that the relevant region is visible.}
	\label{fig2}
\end{figure}

We will therefore make a simplifying large $N$ assumption in the rest of the paper. We will assume that ${m^2\sim N}$ and sometimes write
${m^2=N \tilde m^2}$ \eq{large-mass}; 
this will simplify the calculations, and is far less constraining than $m^2\gg N$. We leave the general case for future work. Repeating the numerical check for the approximation \eq{string-symbol-large-N}, with the above assumption, we find in the Figure \ref{fig2} that the approximation works well for ${1-\cos\vartheta\sim 1/N}$ as desired.
This assumption also changes the small $|x-y|$ behaviour of the symbol and we obtain
 \begin{align}
  \statex{x}{x} \frac 1{\Box+m^2} \state{x}{x}
 &\approx \frac 1N \sum_{l=0}^{\sqrt{N}} \frac{2l+1}{l(l+1)+N\tilde m^2} 
 \sim \frac 1 N \int_0^1 dx \frac{2x}{x^2+\tilde m^2}=\log\left(1+\frac{1}{\tilde m^2}\right) \frac{1}{N}\label{delta-prop-2}
\end{align}
instead of \eq{delta-prop-1}.
Note that the usual UV divergence of the  propagator at coincident points is regularized by the  quantum structure of the geometry, which provides a cutoff at $\L_{\rm NC}$. 
This scale marks the cross-over with the stringy regime of the propagator, which is manifest in \eq{string-prop-two-regimes} and has no classical analog. 
In particular, loop corrections can be computed in a well-defined way in position space:
They consist of standard field theoretic contributions
with cutoff at $L_{\rm NC}$, supplemented by the novel stringy contributions that can be computed effectively using the string symbol for the propagator.

As a first illustration of the use of the above rules, we will rederive the standard results for the one-loop two point function and the UV/IR mixing on the fuzzy sphere \cite{Chu:2001xi}, which was first presented in this formalism in \cite{Steinacker:2016nsc}.

\subsection{One-loop two-point function}\label{sec:one-loop-two-point}

\begin{figure}
	\begin{center}
	\includegraphics[width=0.5\columnwidth]{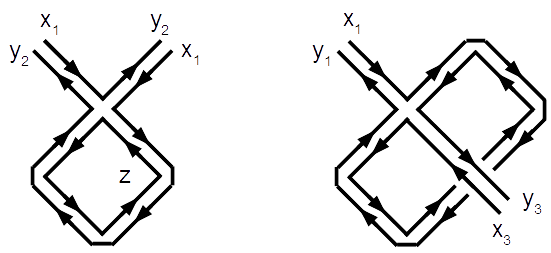}
	\end{center}
	\caption{The two diagrams for the computation of the one-loop two-point function.}
	\label{fig3}
\end{figure}

There are two diagrams contributing to the two point function at one loop, shown in the figure \ref{fig3}.

The first diagram is planar and yields, using the rules (\ref{rules_propagator}) and (\ref{rules_vertex}),
\begin{align}
	\adjincludegraphics[height=2.5cm,valign=c]{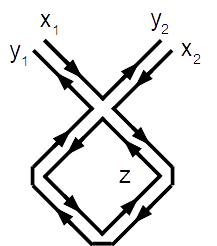}\,=\,&\frac{g}{4!}\frac{4\pi}{N}\frac{N}{4\pi}\left(\frac{N}{4\pi}\right)^4\int dx_4 dy_4 dx_3 dy_3\times\nonumber\\&\times\braket{y_1}{x_2}\braket{y_2}{x_3}\braket{y_3}{x_4}\braket{y_4}{x_1}
	\braket{x_3}{y_4}\braket{x_4}{y_3}\mathcal O_P(\cos\vartheta_{x_4,y_4})=\\=&
	\frac{g}{4!}\braket{y_2}{x_1}\braket{y_1}{x_2}\frac{N}{4\pi}\int dx_4 \mathcal O_P(\cos\vartheta_{x_4,x_1})\ .
	\label{planar-1-loop}
\end{align}
Note how the factors $\braket{y_i}{x_j}$ for the internal labels integrated out trivially using \eq{complete-feynman}, leading to a single internal loop variable $x_4$ with a single factor $\frac{N}{4\pi}$, as advertised before. In the last line, we recognize the reduced ribbon Feynman rules as stated above.

Let us compute the final loop integral explicitly:
\begin{align}
	\adjincludegraphics[height=2.5cm,valign=c]{1loop1}\,=\,&\frac{g}{4!}\braket{y_2}{x_1}\braket{y_1}{x_2}
	\frac{N}{2}\int_0^\pi d\vartheta\,\sin \vartheta\, \mathcal O_P(\cos\vartheta)=	\\
	=& \frac{g}{4!}\braket{y_2}{x_1}\braket{y_1}{x_2}\frac{N}{2}\left(\int_{-1}^{1-\frac{\Delta}{N}}du\, \mathcal O_P(u)+\int_{1-\frac{\Delta}{N}}^1 du\,\mathcal O_P(u) \right)\label{oneloop:P}
\end{align}
where $u=\cos\vartheta$. We have introduced a parameter $\Delta$, which controls the minimal distance of the two points such that the approximation \eq{string-symbol-large-N} is valid. 
Adopting the large mass approximation \eq{large-mass},
we can use this approximation in the first term, while the 
second term can be neglected. More precisely,
we take the maximum value \eq{delta-prop-2} over the whole range of the second integration, overestimating its contribution. Then
\begin{align}
	\adjincludegraphics[height=2.5cm,valign=c]{1loop1}&\approx 
	\frac{g}{4!}\braket{y_2}{x_1}\braket{y_1}{x_2}\frac{N}{2} \left(
	\int_{-1}^{1-\frac{\Delta}{N}} du \frac{1}{\frac{N^2}{4}2(1-u)+m^2}+\frac{\Delta}{N}\frac{1}{N}\right)\\&\approx 
	\frac{g}{4!}\braket{y_2}{x_1}\braket{y_1}{x_2}\frac{N}{2} \left(\frac{2}{N^2}\left( \log N-\log (\tilde m^2+\Delta/2)\right)+\frac{\Delta}{N}\frac{1}{N}\right)\\&\approx \frac{g}{4!}\braket{y_2}{x_1}\braket{y_1}{x_2}\frac{1}{N}\log N
  	=\frac{g}{4!}\braket{y_2}{x_1}\braket{y_1}{x_2}\frac{1}{N}m_N^2\ .\label{mass-renormalization-fromula}
\end{align}
where we have used \eq{string-symbol-large-N}. 
We have dropped the second term as it is subleading in the large $N$ limit. This correctly reproduces the one-loop mass renormalization computed using the standard approach
\begin{align}
m_N^2=\sum_{l=0}^{N-1} \frac{2l+1}{l(l+1)+m^2}\approx \log \frac{N^2}{m^2}
\end{align}
for the ${m^2\sim N}$ case. Note that this assumption was crucial in obtaining \eq{mass-renormalization-fromula}. As we have discussed before, for ${m^2\sim O(1)}$ the approximation of the propagator works only up to $1-\Delta/\sqrt N$ and analogous calculation shows that the second term would dominate the final result. However, $u\to 1$ is the IR regime  where the model becomes commutative, and the IR contribution could be evaluated using commutative methods if desired.

It is interesting to point out that if we did not bother with the validity of the approximation (\ref{string-symbol-large-N}) at all and took the upper limit of the integral (\ref{oneloop:P}) to be $1$, as was done in \cite{Steinacker:2016nsc}, we would have obtained the same result for any value of $m^2$. This suggests that our approach is very conservative and the approximation can be pushed much further. We however leave this for future research.

To compute the contribution from the second non-planar diagram, we again use the Feynman rules to obtain
\begin{align}
	\adjincludegraphics[height=2.5cm,valign=c]{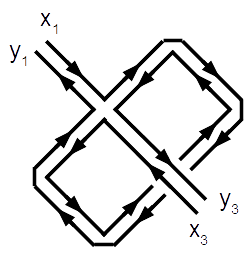}=&
	\frac{g}{4!}\frac{4\pi}{N}\frac{N}{4\pi}\left(\frac{N}{4\pi}\right)^4\int dx_4
	dy_4 dx_2 dy_2\times\nonumber\\&\times\braket{y_1}{x_2}\braket{y_2}{x_3}\braket{y_3}{x_4}\braket{y_4}{x_1}
	\braket{x_2}{y_4}\braket{x_4}{y_2}\mathcal O_P(\cos\vartheta_{x_4,y_4})=\\=&
	\frac{g}{4!}\braket{y_1}{x_1}\braket{y_3}{x_3}\mathcal O_P(\cos\vartheta_{x_1,y_3})\approx \frac{g}{4!}\braket{y_1}{x_1}\braket{y_3}{x_3} \frac{1}{\frac{N^2}{4}|x_1-y_3|^2+m^2}\ .\label{oneloop:NP}
\end{align}
As expected, this diagram has no internal variables to sum over. We also see that the non-planar structure of the diagram is manifest in the different structure of the contractions of the external labels of the diagram.

Together, these two diagrams yield a contribution to the one loop effective action, with graph combinatorial factors taken into account:
\begin{align}
	\frac{g}{3}\left(\frac{N}{4\pi}\right)^4\int & dx_1dy_1dx_2dy_2\,\phi(x_1,y_1)\phi(x_2,y_2)\times\nn\\ &\times\bigg[\braket{y_2}{x_1}\braket{y_1}{x_2}
 \frac{m_N^2}{N}+
 \frac{1}{2}\braket{y_2}{x_2}\braket{y_1}{x_1}\frac{1}{\frac{N^2}{4}|x_1-y_2|^2+m^2}\bigg]\ .
 \label{1-loop-action-general}
\end{align}
Let us point out that the first terms is the single trace quadratic term $\tr(\Phi^2)$ in the matrix model formulation of the fuzzy field theory. This is straightforwardly seen by using $\phi(x,y)=\bra x \Phi \ket y$. Similarly, the second is reminiscent of the multitrace terms like $(\tr\,\Phi)^2$.

When we consider only the local regime of the theory and the field of the form $\phi(x,y)=\phi(x,x)\braket{x}{y}$, denote $\phi(x,x)=\phi(x)$ and use \eq{def:deltaN}, we obtain
\begin{align}
	\frac{g}{3}\frac{1}{4\pi}\int dx\,\phi(x)^2 m_N^2+\frac{g}{6}\left(\frac{N}{4\pi}\right)^2\int dx\, dy\, \phi(x)\phi(y)\frac{1}{\frac{N^2}{4}|x-y|^2+m^2}\ .\label{fuzzy:effaction}
\end{align}
It was observed in \cite{Steinacker:2016nsc} that these two terms reproduce exactly the standard calculation 
in momentum basis \cite{Minwalla:1999px}, which involves an oscillatory integral. It was also noted that the first term is of the same form as the mass term in the bare action (\ref{NCFT-bare-action}) and is thus a mass renormalization. The second term is however very different and yields a nonlocal contribution responsible for  features such as UV/IR mixing.

It is worth writing down the full effective action up to one loop, which is the sum of \eq{fuzzy-action} and \eq{1-loop-action-general}. Restricting to the 
low-energy regime of local functions, this has the form 
\begin{align}
S_{\rm eff}\, =\,& \int dx\, \phi(x)\frac 1 2 (\Box + m^2)\phi(x) +\int dx\, \frac{g}{4!}\, \phi(x)^4\nonumber\\&
 +	\frac{g}{3}\frac{1}{4\pi}\int dx\,\phi(x)^2 m_N^2+\frac{g}{6}\left(\frac{N}{4\pi}\right)^2\int dx\, dy\, \phi(x)\phi(y)\frac{1}{\frac{N^2}{4}|x-y|^2+m^2}\ .
\label{2-loop-eff-action-local}
\end{align}

\subsection{Two-loop effective potential}

In this section, we will reproduce the results of \cite{Huang:2002}, where the contribution of two-loop diagram to the effective action has been computed. This will demonstrate how the presented Feynman rules lead straightforwardly to results, which are quite lengthy to obtain using the standard group-theoretical methods.

\begin{figure}
	\begin{center}
	\includegraphics[scale=0.4]{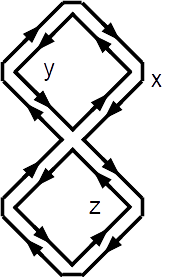}\hskip1cm
	\includegraphics[scale=0.4]{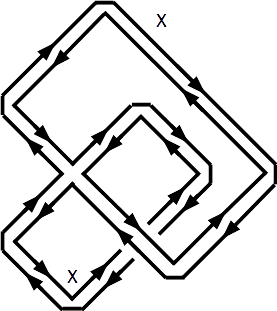}\hskip1cm
	\end{center}
	\caption{The two diagrams for the computation of the two-loop two effective action.}
	\label{fig4}
\end{figure}

The two  diagrams with a quartic vertex that contribute to the effective potential at two-loop order are shown in the figure \ref{fig4}. Using the Feynman rules (\ref{rules_propagator},\ref{rules_vertex}) we can straightforwardly write down the two contributions
\begin{align}
	\adjincludegraphics[height=2.5cm,valign=c]{P}=& 
	\frac{g}{4!}\frac{4\pi}{N}\left(\frac{N}{4\pi}\right)^2\left(\frac{N}{4\pi}\right)^3\int dx\,dy\,dz\,\mathcal O_P(\cos\vartheta_{xz})\mathcal O_P(\cos\vartheta_{xy})\ ,\\
	\adjincludegraphics[height=2.5cm,valign=c]{NP}=& \frac{g}{4!}\frac{4\pi}{N}\left(\frac{N}{4\pi}\right)^2\left(\frac{N}{4\pi}\right)\int dx\,\mathcal O_P(1)\mathcal O_P(1)\ ,
\end{align}
since the label running in the propagators of the second diagram is the same on both sides.

There are several ways to approach the above integrals. For the planar contribution, we can use the approximation \eq{string-symbol-large-N}
and to compare with previous results \cite{Huang:2002} in the $m^2\gg N^2$ limit, we expand in large $m^2$ to obtain
\begin{align}
	\adjincludegraphics[height=2.5cm,valign=c]{P}\approx& 
	\frac{g}{4!}\left(\frac{N}{4\pi}\right)^4 4\pi\left[2\pi \int_{-1}^1 du\frac{1}{\frac{N^2}{2}(1-u)+m^2}\right]^{2}=\nn \\ =&
	\frac{g}{4!}\frac{N^4}{4\pi} \left(
	\frac{1}{m^4}-\frac{N^2}{m^6} +O\left(\frac{N^8}{m^8}\right)\right)\ .
	\label{twoloopP:geometric}
\end{align}
Or we could use the expression \eq{f-Box-expect}, which yields
\begin{align}
	\adjincludegraphics[height=2.5cm,valign=c]{P}\approx& 
	\frac{g}{4!}\frac{N^2}{4\pi}\left[\frac{1}{(4\pi)^2}\int dz \sum_{k,l}\frac{(2l+1)(2k+1)}{l(l+1)+m^2}P_k\left(1-\frac{2l^2}{N^2}\right)P_k(\cos\vartheta)c_k^{-2}\right]^{2}=\nn\\
	=&\frac{g}{4!}\frac{N^2}{4\pi}\left[ \frac{1}{N}\sum_{l}\frac{2l+1}{l(l+1)+m^2}\right]^{2}=
	\frac{g}{4!}\frac{1}{4\pi}\left(\frac{N^4}{m^4}-\frac{N^4(N^2-1)}{m^6} +O\left(\frac{N^8}{m^8}\right)\right)\ ,
\end{align}
where we have used $\int dz P_k(\cos\vartheta)=4\pi\delta_{k,0}$. This is in an agreement with the standard results, with \cite{Huang:2002} and in the large $N$ limit also with \eq{twoloopP:geometric}.

For the non-planar contribution, using \eq{symbol-propagator-local-full} we obtain
\begin{align}
	\adjincludegraphics[height=2.5cm,valign=c]{NP}\approx& \frac{g}{4!}\frac{N^2}{4\pi}\left( \frac{\log N}{N}\right)^2\ ,
\end{align}
where the mass drops out in the large $N$ limit. To compare with the large $m^2$ result, we again use \eq{f-Box-expect}
\begin{align}
	\adjincludegraphics[height=2.5cm,valign=c]{NP}\approx& \frac{g}{4!}\frac{N^2}{4\pi}
	\left[\frac{1}{4\pi N}\sum_{k,l}\frac{(2l+1)(2k+1)}{l(l+1)+m^2}P_k\left(1-\frac{2l^2}{N^2}\right) c_k^{-2}\right]^2=\\
	=&\frac{g}{4!}\frac{N^2}{4\pi}
	\left[\frac{1}{4\pi N}\sum_{k}c_k^{-2}(2k+1)N^2\int_0^1du\, \frac{P_k\left(1-2u\right)}{N^2 u+m^2} \right]^2\ .
\end{align}
We expand the integral in powers of $1/m^2$. In the leading order, only the $k=0$ terms survives, in the next order first two terms survive, however their contributions cancel and we have
\begin{align}
	\adjincludegraphics[height=2.5cm,valign=c]{NP}\approx&\frac{g}{4!}\frac{N^2}{4\pi}\left(\frac{1}{m^4}+O\left(\frac{N^6}{m^6}\right)\right)\ .
\end{align}

We have thus demonstrated that the approach we have presented in the section \ref{sec:one-loop-two-point} works for the calculations necessary to obtain the effective potential and could be in principle used to generalize the results of \cite{Huang:2002} and obtain a more precise analysis of the effective action. Especially interesting is the case of nonzero background $\phi_0$ corresponding to a non-uniform order phase unique to noncommutative field theories. This case requires also cubic diagrams and we leave this for the future and proceed with analysis of the two-loop contribution to the two-point function.

\subsection{Two-loop two-point function}

In this section, we will calculate contributions for few of the diagrams contributing to the two-point function at the two-loop order. We will not give the complete derivation of all the diagrams and the the complete two-point function, but only demonstrate the interesting new concepts compared to the one-loop case.

\begin{figure}
	\begin{center}
	\includegraphics[width=1\textwidth]{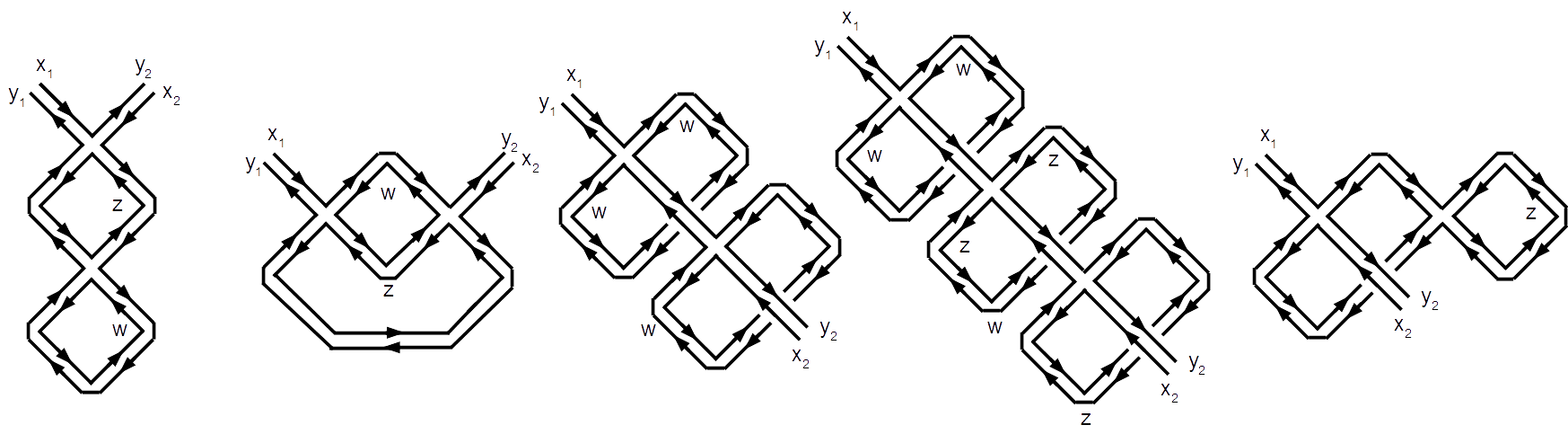}
	\end{center}
	\caption{Several diagrams build using the Feynman rules (\ref{rules_propagator},\ref{rules_vertex}).}
	\label{fig5}
\end{figure}

The diagrams we will consider are shown in the figure \ref{fig5}. To conclude our discussion, let us write down the contributions due to these diagrams and comment on their structure
\begin{align}
	\adjincludegraphics[height=2.5cm,valign=c]{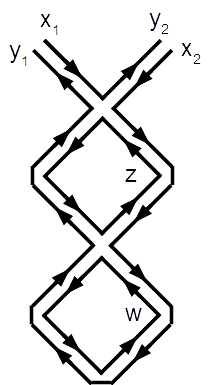}\approx&
	\braket{y_2}{x_1}\braket{y_1}{x_2}
	\left(\frac{g}{4!}\right)^2
	\frac{N}{4\pi}
	\left[\frac{N}{4\pi}\int dw\, \mathcal O_P(\cos \vartheta_{wy_1})\right]
	\left[\frac{N}{4\pi}\int dz\, \mathcal O_P(\cos \vartheta_{zy_1})^2\right]\\
	\adjincludegraphics[height=1.8cm,valign=c]{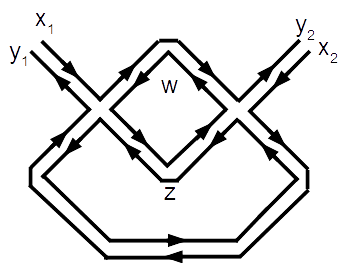}\approx&
	\braket{y_2}{x_1}\braket{y_1}{x_2}
	\left(\frac{g}{4!}\right)^2
	\frac{N}{4\pi}
	\left[\left(\frac{N}{4\pi}\right)^2\int dz dw \,
	\mathcal O_P(\cos \vartheta_{x_1w})
	\mathcal O_P(\cos \vartheta_{wz})
	\mathcal O_P(\cos \vartheta_{zy_1})\right]
\end{align}
These two diagrams are a part of a natural continuation of the standard renormalization process at two-loops. Both are planar and contribute to the ${\phi(x_1,y_1)\phi(x_2,y_2)}$ part of the effective action, but the second diagram introduces a more complicated relation among the labels. In the local regime the second diagram contributes to the effective action with the following term (not taking into account any combinatorial factors)
\begin{align}
    \left(\frac{N}{4\pi}\right)^2\left(\frac{g}{4!}\right)^2\int dx\,\phi(x)^2\left[\left(\frac{N}{4\pi}\right)^2\int dz dw \,
	\mathcal O_P(\cos \vartheta_{xw})
	\mathcal O_P(\cos \vartheta_{wz})
	\mathcal O_P(\cos \vartheta_{zx})\right] \ .
\end{align}

\begin{figure}
	\begin{center}
	\includegraphics[width=0.6\textwidth]{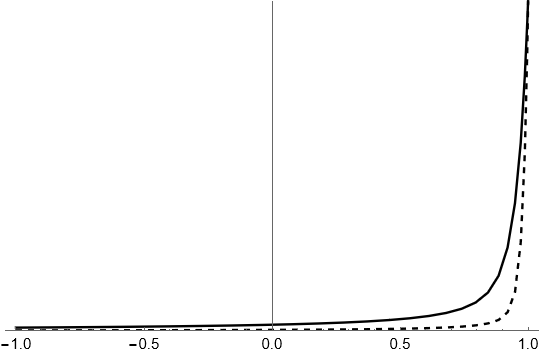}
	\end{center}
	\caption{Comparison of the diagrams \eq{nonloc-iterate} - solid, and \eq{oneloop:NP} - dashed. We show the contribution without the scalar product of the external coherent states and the $g/4!$ factors. The horizontal axis shows $u=\cos\vartheta$ again and we rescale the expressions such that they have the same value for $\vartheta=0$. The plot is for the value $N=120$.}
	\label{fig6}
\end{figure}

Now let us proceed to the nonplanar diagrams.
\begin{align}
	\adjincludegraphics[height=2cm,valign=c]{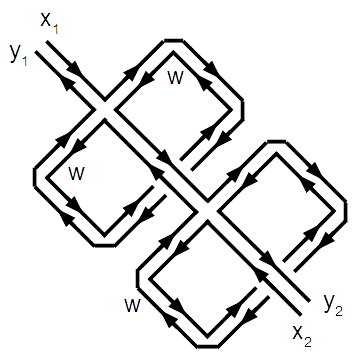}\approx& 
	\braket{y_1}{x_1}\braket{x_2}{y_2}
	\left(\frac{g}{4!}\right)^2
	\frac{N}{4\pi}
	\left[\frac{N}{4\pi}
	\int dw\,\mathcal O_P(\cos \vartheta_{x_1w})\mathcal O_P(\cos \vartheta_{wy_1})\right]
	\mathcal O_P(1)
\label{nonloc-iterate}\\
	\adjincludegraphics[height=2.5cm,valign=c]{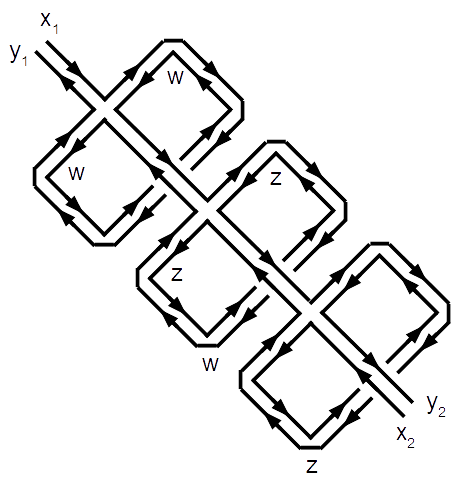}\approx& 
	\braket{y_1}{x_1}\braket{x_2}{y_2}
	\left(\frac{g}{4!}\right)^3 
	\left(\frac{N}{4\pi}\right)^2\mathcal O_P(1)^2\times\nn\\&\times
	\left[\left(\frac{N}{4\pi}\right)^2\int dz dw\,
	\mathcal O_P(\cos \vartheta_{x_1w})
	\mathcal O_P(\cos \vartheta_{wz})
	\mathcal O_P(\cos \vartheta_{zy_1})
	\right]
\label{nonloc-another}
\end{align}
Diagrams like these are higher-order generalizations of the nonplanar diagram \eq{oneloop:NP}.
In particular, the diagram \eq{nonloc-iterate} (and its higher iterations like \eq{nonloc-another}) are responsible for the pathological behavior of noncommutative field theory \cite{Minwalla:1999px}.
The integral is  convergent in 2 dimensions, 
but leads to a novel contribution to the effective action which is even more non-local than the 1-loop contribution \eq{2-loop-eff-action-local}. This is demonstrated in the figure \ref{fig6}, where the contribution of the diagrams \eq{nonloc-iterate} and \eq{oneloop:NP} are compared.
Generalized to 4 dimensions, the integral is 
divergent on non-compact spaces, and
strongly non-local  on compact spaces.

\begin{align}
	\adjincludegraphics[height=2cm,valign=c]{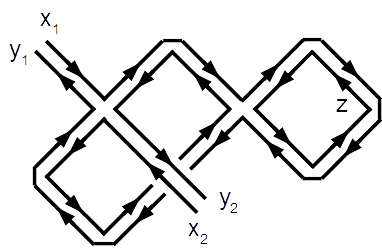}\approx& 
	\braket{y_1}{x_1}\braket{x_2}{y_2}
	\left(\frac{g}{4!}\right)^2
	\frac{N}{4\pi}
	\left[\frac{N}{4\pi}\int dz\, \mathcal O_P(\cos\vartheta_{zx_1})\right]
	\mathcal O_P(\cos\vartheta_{x_1 y_1})^2
\end{align}
This final diagram shows a similar behaviour as the one-loop diagram (\ref{oneloop:NP}), introducing an explicit coupling between $\phi(x)$ and $\phi(y)$ in the effective action (\ref{fuzzy:effaction}) with a different and more non-local dependence $\sim(N^2|x_1-y_1|^2/4+m^2)^{-2}$.

\section{Discussion}

We have discussed in detail the string representation of the propagator on the fuzzy sphere. It provides an explicit separation of the degrees of freedom in non-commutative field theory into a low-energy semi-classical and almost local regime, and a high-energy, stringy and non-local regime. While the low-energy regime behaves approximately as in standard field theory, the stringy regime displays a very different behavior, which can effectively captured by the methods developed here. It turns out that the stringy regime in noncommutative field theory is very different and in fact much simpler than in ordinary field theory, described by ribbon graphs which are manifestly local and diagonal. The stringy regime turns out to dominate for sufficiently large mass ${m^2 = O(N)}$, which should allow to obtain  analytical control over phase transitions of various models; this is postponed for future work. 
  
An important application of the present framework is to compute the quantum effective action of noncommutative field theory and  matrix models directly in position space, thereby avoiding 
oscillatory integrals.
We formulate the appropriate Feynman rules for the case of $S^2_N$, which straightforwardly generalize to more general quantum spaces.
A crucial and novel feature is that the identification  \eq{wavepacket} of short string modes as localized wave-packets allows to evaluate the effective action directly in position space,
even on non-trivial backgrounds.
This was recently used to obtain the Einstein-Hilbert action from the one-loop effective action of the maximally supersymmetric IKKT or IIB matrix model \cite{Steinacker:2021yxt}, where the stringy regime is suppressed by SUSY.

\section*{Acknowledgments}
The work of HS is supported by the Austrian Science Fund (FWF) grant P32086, the work of JT is supported by VEGA 1/0703/20 grant \emph{Quantum structure of spacetime}.

\appendix

\section{Appendix: useful identities}

We collect some useful formulas here. 

The spherical harmonics addition theorem states that
\begin{align}
 \sum_m  Y^{l}_{m}(x) Y^{l*}_{m}(y) 
 = \frac{2l+1}{4\pi} P_l(\cos\vartheta)\ .
 \label{spher-add-thm}
\end{align}
For the 6j symbols, we will need the following sums
\begin{align}
\sum_{k=0}^{2\a}(2k+1) k(k+1) (-1)^{k+2\a} \left\{
\begin{array}{lll}
 0 & \a & \a \\
 k & \a & \a
         \end{array}\right\}
           &= \frac 12 N (N^2-1)\ , \nn\\
 \sum_{k=0}^{2\a}(2k+1) k(k+1) (-1)^{k+2\a} \left\{
\begin{array}{lll}
 1 & \a & \a \\
 k & \a & \a
         \end{array}\right\}
          &=  \frac{1}{6} N (N^2-1)
\label{6J-sums-1}
\end{align}
and
\begin{align}
\sum_{k=0}^{2\a}(2k+1) (k(k+1))^2 (-1)^{k+2\a} \left\{
\begin{array}{lll}
 0 & \a & \a \\
 k & \a & \a
         \end{array}\right\}
          &= \frac{1}{3} N \left(N^2-1\right)^2\ , \nn\\
 \sum_{k=0}^{2\a}(2k+1) (k(k+1))^2 (-1)^{k+2\a} \left\{
\begin{array}{lll}
 1 & \a & \a \\
 k & \a & \a
         \end{array}\right\}
          &= \frac{1}{6} N (N^2-1)\left(N^2-2\right)\ ,\nn\\
 \sum_{k=0}^{2\a}(2k+1) (k(k+1))^2 (-1)^{k+2\a} \left\{
\begin{array}{lll}
 2 & \a & \a \\
 k & \a & \a
         \end{array}\right\}
          &= \frac{1}{30} N (N^2-1) (N^2-4)\ .
\label{6J-sums-2}
\end{align}

\section{Appendix: asymptotic form for $c_l$}
\label{sec:asymptotic}

Explicitly,
\begin{align}
 c_0^2 
  &= \frac{N}{4\pi}\ ,  \nn\\
  c_1^2 
   &= \frac{1}{4\pi}\frac{(N + 1)N}{N-1} \ ,
   \label{c-i-explicit}
\end{align}
and
\begin{align}
 \hat Y^0_0 &= \frac 1{\sqrt{N}} \one\ ,  & Y^0_0(x) &= \frac{1}{\sqrt{4\pi}} \ , \nn\\
 \hat Y^1_m &= \sqrt{\frac{3}{N}}\,X_m\ ,  & Y^1_m(x) &= \sqrt{\frac{3}{4\pi}} x_m , \ \ m=0,\pm 1 \ .
\end{align}
For $l=0$, \eq{hatY-Y-relation} reproduces the completeness relation 
\begin{align}
\one = \frac{N}{4\pi} \int dx\, \ket x \bra x \ .
\end{align}
For $l=1$, this gives
\begin{align}
 X^a 
 = \sqrt{\frac{N+1}{N-1}} \frac{N}{4\pi} \int_{S^2} dx\, \ket x \bra x x^a \ .
 \label{X-rep-coherent}
\end{align}

\paragraph{Asymptotic behavior of $c_l$.}

An  analytic form for the asymptotic behavior of $c_l$ can be obtained as follows
\begin{align}
  2\ln c_l 
   &\approx \ln(N/4\pi) 
   + \sum_{m=1}^{l} \ln\Big(\frac{1+m/N}{1-m/N}\Big)  \nn\\
   &\sim \ln(N/4\pi) 
   + N\int_{0}^{l/N} \ln\Big(\frac{1+u}{1-u}\Big)du  \nn\\
 &= \ln(N/4\pi) 
   + N(x+1) \ln \left(\frac{x+1}{1-x}\right)+2N \ln (1-x)|_{x=\frac{l}{N}}
\end{align}
where 
\begin{align}
 x=\frac{l}{N}
\end{align}
Taking the leading order small $x$ behaviour of this expression, we find
\begin{align}
\boxed{\
 c_l^2 \sim \frac{N}{4\pi} e^{N x^2}, 
 \qquad x = \frac{l}{N}
\ }
\label{c-n-asymptotic-app}
\end{align}
Numerically, this is seen to be a good fit for the entire range. It underestimates the expression slightly for larger values of $l$, which however only strengthens the argument that $c_l^{-2}$ cut-off sums over $l$ at $l\sim\sqrt N$.

\section{Appendix: proof of Lemma \ref{lem:cross}}
\label{sec:app-lemma}

To prove Lemma \ref{lem:cross} for invariant operators $\cO$, 
we note that the crossing operation $\cC$  \eq{C-crossed} preserves 
$SO(3)_{\rm diag}$ - invariant operators. Therefore there is a matrix $A^{lk}$ such that
\begin{align}
 \sum_m\hat Y^l_m \otimes \hat Y^{l\dagger}_{m}
   &= \sum_{m,r,r',s,s'}  (-1)^m (2l+1)\begin{pmatrix}
               l & \a & \a \\
               m & r & -s
              \end{pmatrix}
             \begin{pmatrix}
               l & \a & \a \\
               -m & r' & -s'
              \end{pmatrix}
              |r\rangle\langle s|\otimes |r'\rangle\langle s'| \nn\\
&= \sum_{mrr'ss'}(-1)^m (2l+1) \begin{pmatrix}
               l & \a & \a \\
               m & r & -s
              \end{pmatrix}
              \begin{pmatrix}
               l & \a & \a \\
               -m & r' & -s'
              \end{pmatrix}
              \cC(|r\rangle\langle s'|\otimes |r'\rangle\langle s|) \nn\\
 &\stackrel{!}{=} \sum_{k} A^{l k} \cC(\sum_n\hat Y^k_n \otimes \hat Y^{k\dagger}_{n})\ ,
\end{align}
with $A^2 = \one$
using \eq{hat-Y-CG-1}. Note that both sides are invariant under $SO(3)_{\rm diag}$.
The $A^{lk}$ can be extracted by 
taking the inner product with 
\begin{align}
 \cC(\sum_n\hat Y^k_n \otimes \hat Y^{k\dagger}_{n}) = 
(-1)^{n+r+r'+s+s'}(2k+1)\sum\limits_{nrr'ss'} \begin{pmatrix}
               k & \a & \a \\
               n & r & -s'
              \end{pmatrix}
              \begin{pmatrix}
               k & \a & \a \\
               -n & r' & -s
              \end{pmatrix}
              |r\rangle\langle s|\otimes |r'\rangle\langle s'|  \nn  
\end{align}
Dropping an overall factor $(2k+1)$ and 
and using the above orthogonality relation, this gives
\begin{align}\small
 A^{l k} &= 
 (2l+1)\sum_{mrr'ss'} (-1)^{m-n+r+r'+s+s'}\begin{pmatrix}
               l & \a & \a \\
               m & r & -s
              \end{pmatrix}
              \begin{pmatrix}
               l & \a & \a \\
               -m & r' & -s'
              \end{pmatrix} \begin{pmatrix}
               k & \a & \a \\
               n & r & -s'
              \end{pmatrix}
              \begin{pmatrix}
               k & \a & \a \\
               -n & r' & -s
              \end{pmatrix} \nn\\
&= (2l+1)(-1)^{l+k+2\a} \left\{
\begin{array}{lll}
 k & \a & \a \\
 l & \a & \a
         \end{array}
     \right\}\ ,
\label{Akl-explicit}
\end{align}
using the definition of the 6j symbols
\begin{align}
(-1)^{2\a}\left\{
\begin{array}{lll}
k & \a & \a \\
l & \a & \a
         \end{array}
     \right\} =
 \sum_{m,n,r,r',s,s'}& (-1)^{l+k-m-n+r+r'+s+s'}\begin{pmatrix}
               l & \a & \a \\
               m & r & -s
              \end{pmatrix}
              \begin{pmatrix}
               l & \a & \a \\
               -m & r' & -s'
              \end{pmatrix}\times\nn\\&\times
\begin{pmatrix}
               k & \a & \a \\
               -n & -r & s'
              \end{pmatrix}
              \begin{pmatrix}
               k & \a & \a \\
               n & -r' & s
              \end{pmatrix}
\end{align}
and their standard symmetry properties.
The property $\cC^2 = \one$, i.e. $A^2 = \one$, follows from  
the orthogonality relation for the 6j symbols.

This completes the proof of Lemma \ref{lem:cross}.

\end{document}